\documentclass[%
preprint,
 amsmath,amssymb,
 aps,
]{revtex4-2}
\usepackage{appendix}
\usepackage{yfonts}
\usepackage{xcolor}
\usepackage{xcite}
\usepackage{braket}
\usepackage{graphicx}
\usepackage{dcolumn}
\usepackage{bm}

\usepackage[english]{babel}
\usepackage{blindtext}
\begin{document}

\preprint{APS/123-QED}

\title{
Entropy in Loop Quantum Cosmology}

\author{Alejandro Corichi}
 \email{corichi@matmor.unam.mx}
 \affiliation{%
Centro de Ciencias Matemáticas, Universidad Nacional Autónoma de México, UNAM-Campus Morelia, A. Postal 61-3, Morelia, Michoacán C.P. 58090, México.}%
\author{Omar Gallegos}
 \email{omar.gallegos@iem.cfmac.csic.es}
\affiliation{%
Centro de Ciencias Matemáticas, Universidad Nacional Autónoma de México, UNAM-Campus Morelia, A. Postal 61-3, Morelia, Michoacán C.P. 58090, México.}
\affiliation{%
Instituto de Estructura de la Materia, IEM-CSIC, C/ Serrano 121, 28006 Madrid, Spain}

\date{\today}

\begin{abstract}
 The Generalized First Law (GFL) and the Generalized Second Law (GSL) of thermodynamics are studied for cosmological scenarios with spatial curvature through an apparent horizon. We focus on effective and alternative cosmic systems motivated by quantum cosmological models, where the entropy is considered a function of the apparent area, transforming the effective cosmological model into the standard form in cosmology. The general conditions for the validity of the GSL are analyzed for entropy as a general function of area and logarithmic corrections to the usual Black Hole entropy. The Weak Energy Condition (WEC) and the Strong Energy Condition (SEC) are implemented for the matter entropy part. In particular, we study the GFL and the regions where the GSL is valid for effective Loop Quantum Cosmology (LQC) models with spatial curvature $k=0,\pm 1$, taking every possible value of the logarithmic contributing factor for the entropy analysis. In addition, in order to solve some violations of the GSL, we explore the possibility of admitting negative absolute temperatures (NAT) in our system, where the validity conditions for an extended generalized second law (EGSL) are studied, and the time arrow from the second law is discussed for the LQC models.

\end{abstract}

\maketitle


\color{black} 

\section{Introduction}
\label{section:introduction.}
Entropy is a fundamental concept in thermodynamics, emerging from the first law of thermodynamics through the exchange of energy and the work performed by a system. In statistical mechanics, entropy is related to the counting of microstates that correspond to a macrostate. Although most equations and laws of physics are invariant under time reversal, entropy is unique in providing a direction of time through the second law of thermodynamics, which states that the entropy of a closed system can only increase or remain constant. As an empirical theory, thermodynamics plays a central role across all fields of physics.

The relationship between geometry and thermodynamics is well established in black hole (BH) physics \cite{Bekenstein:1973ur,Hawking:1975vcx}, where it was shown that these objects radiate and that temperature and entropy are associated with geometric quantities at the horizon. In contrast, cosmological thermodynamics has been less extensively explored. Nonetheless, several approaches have been proposed to extend these ideas to general space-times \cite{Jacobson:1995ab,Eling:2006aw}. The thermodynamic formulation of black hole physics was among the first frameworks to suggest a deep connection between quantum mechanics (QM) and general relativity (GR) through thermodynamic laws. This formulation was originally developed within the semiclassical approximation, with the expectation that it would be improved in a full quantum gravity theory.

Loop Quantum Gravity (LQG) is a non-perturbative, background-independent approach to quantum gravity that aims to unify QM and GR \cite{Thiemann:2007pyv,Rovelli:2004tv}. Classical singularities in black holes and cosmology are resolved in symmetry-reduced models through LQG techniques. Although there is no notion of space or time in the full theory, the entropy calculation based on microstate counting has been developed, yielding quantum corrections to black hole entropy and fixing the Barbero–Immirzi parameter \cite{Ashtekar:1997yu,Rovelli:1996dv,Meissner:2004ju,Ghosh:2004rq,Ghosh:2006ph,Agullo:2010zz-entropyBH,Engle:2009vc,Domagala:2004jt}.

Loop Quantum Cosmology (LQC) applies these methods to cosmological settings as a testbed for quantum gravity effects, with loop-quantization implemented in a symmetry-reduced model \cite{Rovelli:2004tv,Thiemann:2007pyv}. In LQC, classical singularities are replaced by a quantum bounce connecting two cosmological branches \cite{Bojowald:2001xe,Bojowald:2005epg,Bojowald:2011ku, Ashtekar:2006wn, Ashtekar:2007em}. From black hole and cosmological models, we can associate geometry with thermodynamic quantities. Although LQC equations are invariant under time reversal, we can ask what the behavior of entropy is `before' and `after' the bounce? Because entropy is a function of horizon area, a branch of the Universe contracts until the quantum bounce, which is the minimum area, and then the universe expands. However, it will not be necessary to go beyond the bounce to find the regions where the thermodynamic laws fail.

Thermodynamics in LQC has been studied to address the covariant entropy conjecture \cite{Bousso:1999xy,Ashtekar-Ewing:2008gn} using the particle horizon, where it is possible to find a finite value of the ratio $S/A$ throughout the universe's evolution. On the other hand, the apparent horizon was used to explore the first law for the gravitational part applied to two different effective flat models in LQC using entropy calculus without logarithmic corrections \cite{Li-Thermo:2008tc,Zhang-Thermo:2021umq}. In addition, the Generalized Second Law (GSL) is studied for only some logarithmic correction factors in an effective flat LQC \cite{Sadjadi:2012wg}. The GSL is the sum of the change of each entropy contribution of a system. In this case, matter and gravitational entropies. In addition, a different function of the apparent area for entropy is used to explore the GSL validity conditions for the same LQC model \cite{Silva:2023ent,Hossenfelder:2012tc}. Both analyzes of the GSL in LQC suggest that the GSL is not valid close to the quantum bounce.

Throughout this work, we generalize and complete previous investigations of thermodynamics in loop quantum cosmology (LQC) at the apparent horizon. The gravitational part of the first law is formulated by treating entropy as a function of the apparent horizon area, applicable to several quantum cosmological models that can be recast in standard cosmological form. The matter contribution is then included, yielding the generalized first law of cosmology. This matter sector is assumed to consist of ordinary matter satisfying the weak energy condition (WEC), $\rho\geq 0$, and the strong energy condition (SEC), $\rho +3P\geq 0$, while also requiring $\rho+P\geq 0$, where $\rho$ and $P$ are its energy density and pressure, respectively. Furthermore, general conditions for the validity of the generalized second law are derived for arbitrary effective cosmological models and for general entropy–area relations.

As a particular case, we consider logarithmic corrections derived from black hole entropy calculations in LQG \cite{Ashtekar:1997yu,Rovelli:1996dv,Meissner:2004ju,Ghosh:2004rq,Ghosh:2006ph,Agullo:2010zz-entropyBH,Engle:2009vc,Domagala:2004jt}, where the logarithmic correction parameter $\tilde{\alpha}$ modifies the gravitational energy, volume, entropy, and work terms in the GFL. In the limit $\tilde{\alpha}\to 0$, the standard gravitational first law for the effective flat LQC model is recovered \cite{Li-Thermo:2008tc}. We also derive the conditions under which the GSL holds for general effective cosmological models with logarithmic corrections.
In particular, we analyze effective LQC models with flat \cite{Ashtekar:2006wn,Ashtekar:2007em,Ashtekar:2003hd}, hyperbolic \cite{Vandersloot:2006ws,Szulc:2007uk}, and closed \cite{Ashtekar:2006es,Szulc:2006ep} geometries, which can be reformulated in the standard framework of classical cosmology. This reformulation allows us to use standard cosmological tools to study their thermodynamic properties.

The regions $0<\rho<\rho_*/2$ and $\rho_*/2<\rho<\rho_*$ affect the conditions under which GSL is valid. Here, $\rho_*$ is the maximum energy density for each geometry. These regions are associated with changes in the sign of the quantity $\dot{H}-\frac{k}{a^2}$.  Furthermore, we analyze, for each geometry and every possible value of the logarithmic correction factor, that depending on the counting of microstates in BH, this factor could have different values.

The GFL and GSL for the LQC models with spatial curvature distinct from zero have not been previously studied in the literature. However, for the flat case, when we consider the case $\tilde{\alpha}=-\frac{3}{2\rho_c}$, we obtain the results in \cite{Sadjadi:2012wg}. Thus, previous results for entropy in LQC are included as particular cases in our analysis, where we consider an apparent horizon to explore the thermodynamics of effective cosmological models, and we extend this study to closed and open models.

In addition, previous analyzes of the flat LQC model indicate that the GSL is not universally valid \cite{Sadjadi:2012wg,Silva:2023ent}. To address this issue, we extend the framework beyond the $k=0$ case and consider effective LQC models with spatial curvature. We further investigate the possibility of negative absolute temperature (NAT), which is physically admissible under specific conditions \cite{Ramsey-NegativeTemp:1956zz,BALDOVIN20211-NegativeTemp,Corichi:2025tts}, and examine whether the thermodynamic laws remain valid in such regimes.

Motivated by this, we put forward an extension of the second law that incorporates both positive and negative temperature domains. This \textit{extended generalized second law} (EGSL) alleviates certain violations of the standard GSL and extends its regime of validity when NAT configurations are included. Within this formulation, the apparent inconsistencies are resolved, and the framework may be applied more generally beyond quantum cosmology to the study of thermodynamic laws in systems exhibiting negative absolute temperature \cite{Corichi:2025tts}.

Although microstate counting for black holes in LQG is relevant for fixing the value of the Barbero Immirzi parameter, the logarithmic corrections do not affect this result. This is because, for this calculation, it is necessary to obtain the asymptotic limit, which occurs when the area is large; in that limit, the logarithmic corrections are negligible. 

This article is organized as follows. In Section \ref{section:Entropy in FLRW}, we present the standard formalism for constructing the thermodynamic description of a Friedmann–Lemaître–Robertson–Walker (FLRW) spacetime using the apparent horizon. We consider effective and alternative quantum cosmological models, noting that the apparent radius depends on the spacetime geometry but is independent of the specific gravitational theory. Using entropy as a function of the apparent horizon area and assuming the SEC, we derive the generalized first law and generalized second law, and obtain the corresponding conditions for the validity of the GSL.
In Section \ref{section: log corrections eff}, logarithmic corrections to the entropy are introduced, allowing for the formulation of the GFL for general effective cosmological models. The resulting regions of validity of the GSL are also identified. In Section \ref{sec: LQC}, we review effective LQC models with spatial curvature and show how they can be rewritten in a standard cosmological form, leading to a unified description in terms of the curvature parameter $k$, as in standard cosmology.

In Section \ref{subsec: GFL in LQC}, we apply the unified formulation to derive the GFL for effective LQC models. The flat case is recovered as a particular limit, reproducing both the case without logarithmic corrections and the classical result. In Section \ref{subsec: GSL in LQC}, we further analyze the GSL for semiclassical LQC models with $k=0,\pm 1$, assuming the WEC and SEC for the matter sector. We find that the GSL is violated near the quantum bounce.

In Section \ref{Appe: negative T}, we consider the possibility of negative absolute temperature and introduce an extension of the generalized second law, expressed as $\mathrm{sgn}(T)\mathrm{d}S\geq0$, which includes the standard condition $\mathrm{d}S\geq0$ when only $T>0$ is considered. We compare the EGSL and GSL in effective LQC models with spatial curvature, where the novel results that some inconsistencies of the GSL are solved. Finally, in Section \ref{Time arrow}, we discuss the thermodynamic behavior across the quantum bounce and define a time arrow based on time-reversal invariance. Conclusions are presented in Section \ref{section:Conclusion}.

\section{Cosmic Entropy as a function of the apparent area.}
\label{section:Entropy in FLRW}
The FLRW cosmological model, in which isotropy and homogeneity are considered, is described by the line element
\begin{equation}
\label{FLRW metric}
    \mathrm{d}s^2=-N\mathrm{d}t^2+a^2(t)\left(\dfrac{\mathrm{d}r^2}{1-kr^2}+r^2\mathrm{d}\Omega\right),
\end{equation}
where $N$ is the lapse function, $a(t)$ is the scale factor, the 2d line element is denoted by $\mathrm{d}\Omega$, and the spatial curvature $k=-1,0,+1$ corresponds to hyperbolic, flat, and spherical universes, respectively. Now, we assume an energy-momentum tensor $T_{\mu\nu}$ for a perfect fluid
\begin{equation}
\label{energy-momentum tensor perfect fluid}
    T_{\mu\nu}=(\rho+P) U_\mu U_\nu+P g_{\mu \nu}.
\end{equation}
We can find the continuity equation 
\begin{eqnarray}
      \label{FLRW continuity eq}
      \Dot{\rho}+3H(\rho+P)=0,
\end{eqnarray}
where $\Dot{\rho}$ represents the derivative of $\rho$ w.r.t. the cosmic time $t$, $\rho$ is the energy density, and $P$ stands for the matter pressure. Without loss of generality, we use $N=1$ and rewrite the line element as
\begin{equation}
    \label{FLRW metric 2}
     \mathrm{d}s^2=h_{ab}\mathrm{d}x^a\mathrm{d}x^b+R^2\mathrm{d}\Omega,
\end{equation}
where $R=a(t) r$ and the 2-dimensional metric $h_{ab}=\mathrm{diag}(-1,a/(1-kr^2))$.

To analyze the thermodynamics of a cosmological model, it is necessary to introduce a horizon. Conveniently, a dynamical apparent horizon is used, that is, a marginally trapped surface with vanishing expansion which fulfills the following relation $h^{ab}\partial_a R\partial_b R=0$ \cite{Hayward:1998ee, Hayward:1997jp,Cai:2005ra,Faraoni:2015ula, Silva:2023ent}. Therefore, the radius of the cosmic apparent horizon $R_A$ is given by  
\begin{eqnarray}
    \label{apparent radius}
    R_A=\dfrac{1}{\sqrt{H^2+k/a^2}},
\end{eqnarray}
where $H=\frac{\dot{a}}{a}$ is the Hubble parameter. In addition, note that the apparent horizon is only defined when $H^2+k/a^2> 0$ because the WEC is valid. Furthermore, for $k=0$ the apparent radius coincides with the radius of the Hubble horizon $R_H=1/H$. On the other hand, the particle horizon $R_p$ in a time interval between $t_i$ and $t_f$ is defined as 
\begin{equation}
\label{particle horizon}
    R_p=a(t) \int_{t_i}^{t_f} \frac{\mathrm{d} t}{a(t)},
\end{equation}
when $t_f\to\infty$, the particle radius transforms to the radius of the cosmological event horizon $R_E$. For a flat de Sitter Universe, $R_A$ and $R_E$ take the same expression. In \cite{Ashtekar-Ewing:2008gn} the radius of the particle horizon (\ref{particle horizon}) is used to solve the covariant entropy conjecture in the LQC context \cite{Bousso:1999xy}. However, thermodynamics description depends on the choice of the horizon in which we work \cite{Faraoni:2015ula}. For a dynamical space-time, the apparent horizon is a causal horizon, and it is related to thermodynamics through entropy, gravitational energy, and the gravity surface \cite{Hayward:1997jp,Hayward:1998ee,Cai:2005ra}.

It is possible to define the density work $W$ and the energy-supply vector $\psi_a$ as
\begin{align}
    \label{work density}
W=-\dfrac{1}{2}T^{ab}h_{ab},\\
\label{energy-supply vector}
\psi_a=T_a^b \partial_b R+W \partial_a R,
\end{align}
for the FLRW case eqs. (\ref{work density}) and (\ref{energy-supply vector}) take the following form  
\begin{align}
    \label{FLRW work density}
W&=\dfrac{1}{2}\left(\rho-P\right),\\
\label{FLRW energy-supply vector}
\psi_a&=\left(-\frac{1}{2}(\rho+P) H R, \frac{1}{2}(\rho+P) a\right) .
\end{align}
The work density at the apparent horizon is the work done by a change of the apparent radius. In addition, the energy-supply at this horizon is the total energy flow through the apparent horizon
\begin{equation}
    \nabla E_{g}= A_A\psi+W\nabla V_A,
\end{equation}
where $A_A=4\pi R^2$, $V_A=\frac{4\pi}{3}R^3_A$ are the apparent area and apparent volume, respectively. Furthermore, the Misner-Sharp energy $E_{g}$ is defined as the total gravitational energy inside of a sphere with radius $R$ \cite{Misner:1964je,Ashworth:1998uj}  
\begin{equation}
\label{Misner-Sharp energy}
    E_{g}=\dfrac{R}{2G}\left(1-h^{ab}\partial_a R\partial_b R\right).
\end{equation}
If we consider only the Misner-Sharp energy at the apparent horizon $R_A$, then $E_g=R_A/(2G)$.

This formalism is independent of the specific theory of gravity under consideration; it depends only on the background spacetime, which is taken to be FLRW. One may adopt a gravitational theory beyond standard cosmology, such as an effective field theory approach. However, the resulting equations can still be written in the same form as the standard cosmological equations by appropriately defining an effective energy density $\rho_\mathrm{eff}$ and an effective matter pressure $P_\mathrm{eff}$, such that 
\begin{align}
\label{eff friedmann eq}
&H^2=\dfrac{8\pi G}{3}\rho_\mathrm{eff}-\dfrac{k}{a^2},\\
    \label{eff Raychaudhuri eq}
&\frac{\ddot{a}}{a}=\dot{H}+H^2=-\frac{4 \pi G}{3}\left(\rho_{\mathrm{eff}}+3 P_{\mathrm{eff}}\right), \\
\label{eff continuity eq}
&\dot{\rho}_{\mathrm{eff}}+3 H\left(\rho_{\mathrm{eff}}+P_{\mathrm{eff}}\right)=0,\\
\label{eff dot H}
&\dot{H}=-4\pi G(\rho_\mathrm{eff}+P_\mathrm{eff})+\dfrac{k}{a^2},
\end{align}
where the effective Friedmann equation is written in (\ref{eff friedmann eq}), the Raychaudhuri equation reads as in (\ref{eff Raychaudhuri eq}), the continuity equation is given in (\ref{eff continuity eq}) and the equation (\ref{eff dot H}) is obtained using (\ref{eff friedmann eq}) and (\ref{eff continuity eq}). We recover standard cosmological equations if $\rho_\mathrm{eff}\to \rho$ and $P_\mathrm{eff}\to P$. This way to rewrite the effective equations of a theory is convenient to use techniques that have already been developed in general relativity.

Therefore, using (\ref{apparent radius}) and (\ref{eff friedmann eq}) the gravitational energy at the apparent horizon is given by 
\begin{equation}
\label{apparent energy}
    E_{g}=\dfrac{4\pi R_A^3}{3}\rho_\mathrm{eff}=V_A \rho_\mathrm{eff}.
\end{equation}
To examine the thermodynamics of a cosmic model, we associate a temperature $T$ and a gravitational entropy $S_g$, with the apparent horizon. The temperature is related to the surface gravity, while the entropy is taken to be a function of the horizon area. These relations are motivated by black hole thermodynamics in general relativity, where the Bekenstein–Hawking entropy is given by $A/(4\ell^2_p)$, with $A$ the horizon area and $\ell_p=\sqrt{G\hbar}$ the Planck's length \cite{Hawking:1975vcx,Bekenstein:1973ur}. By applying the same ansatz in cosmology, one can recover the standard Friedmann equations \cite{Hayward:1997jp,Hayward:1998ee,Cai:2005ra,Faraoni:2015ula} (Throughout, we work in natural units $G=\hbar=c=1$). 

BH entropy calculus combines, at the semiclassical level, QM and GR. However, with an improved counting of microstates from (loop) quantum gravity, we can find different proposals where the entropy is not only proportional to area but rather a function of the horizon area \cite{Ashtekar:1997yu,Ghosh:2004rq,Ghosh:2006ph,Rovelli:1996dv,Meissner:2004ju,Engle:2009vc,Agullo:2010zz-entropyBH,Hossenfelder:2012tc,Domagala:2004jt}. With this motivation, we consider the general case where the gravitational entropy $S_g$ is a function of the apparent area, which is given by 
\begin{align}
    \label{f(A) entropy}
    S_g&=f\left(\dfrac{A_A}{4}\right),
\end{align}
the surface gravity $\kappa$ is defined as
\begin{equation}
\label{surface gravity}
    \kappa=\frac{1}{2 \sqrt{-h}} \partial_a\left(\sqrt{-h} h^{a b} \partial_b R\right) ,
\end{equation} 
at the cosmic apparent horizon, the surface gravity is written as 
\begin{align}
    \label{FLRW surface gravity}
    \kappa&=-\dfrac{R_A}{2}\left(\Dot{H}+2H^2+\dfrac{k}{a^2}\right),\nonumber\\
    &=-\dfrac{1}{R_A}\left(1-\dfrac{\Dot{R}_A}{2HR_A}\right).
\end{align}
Surface gravity is associated with the attractive or repulsive gravitational interaction. Note that $\kappa$ could take positive, negative, or zero values. However, we only consider $T>0$, or equivalently $T=-\mathrm{sgn}(\Dot{H}+2H^2+\frac{k}{a^2})\frac{\kappa}{2\pi}$. This is due to the fact that $H^2+\frac{k}{a^2}\geq 0$ but $\dot{H}-\frac{k}{a^2} \lessgtr 0$, then $\Dot{H}+2H^2+\dfrac{k}{a^2} \lessgtr0$. Therefore, the temperature that we consider, it is defined as  
\begin{align}
     \label{cosmic temperature}
    T&=\left\vert\dfrac{\kappa}{2\pi}\right\vert.
\end{align}
\subsection{Generalized First Law}
\label{subse: GFL f(A)}
To find the first law of the cosmic thermodynamics, we use (\ref{f(A) entropy}) and (\ref{cosmic temperature}) obtaining 
\begin{align}
\label{TSg}
    \left\vert\dfrac{\kappa}{2\pi}\right\vert\mathrm{d}S_g=\mathrm{sgn}\left(\Dot{H}+2H^2+\frac{k}{a^2}\right)\left(1-\dfrac{\dot{R}_A}{2H R_A}\right)\dot{R}_A f'(A_A)\mathrm{d}t,
\end{align}
where $f'(A_A)=\frac{\mathrm{d}f(\frac{A_A}{4})}{\mathrm{d}\frac{A_A}{4}}$. Then, deriving the gravitational energy (\ref{apparent energy}) and using the continuity expression (\ref{eff continuity eq}), we find 
\begin{equation}
\label{derivative energy g}
    \mathrm{d} E_g=4 \pi \rho_\mathrm{eff} R_A^2 \dot{R}_A \mathrm{d} t-4 \pi R_A^3 H\left(\rho_\mathrm{eff}+P_\mathrm{eff}\right) \mathrm{d} t.
\end{equation}
We can obtain the first law taking into account (\ref{TSg}) and (\ref{derivative energy g}) 
\begin{align}
    \label{f(A) first law}
    \mathrm{D}E_g=T\mathrm{d}S_g+\tilde{W}\mathrm{D}V_A,
\end{align}
where the effective work $\tilde{W}$ is given by 
\begin{align}
    \label{eff work}
\tilde{W}&=\dfrac{1}{2}\left(\rho_\mathrm{eff}-P_\mathrm{eff}\right),
\end{align}
and 
\begin{align}
    \mathrm{D}E_g&=f'(A_A)\mathrm{d}E_g,\\
    \mathrm{D}V_A&=f'(A_A)\mathrm{d}V_A,
\end{align}
to recover the first law in standard cosmology, we need that $f'(A_A)\to 1$, $\rho_\mathrm{eff}\to \rho$ and $P_\mathrm{eff}\to P$ when $A_A$ is large. Now, to get the Generalized First Law (GFL) of thermodynamics, we add a contribution of matter in (\ref{f(A) first law}), which obeys the SEC
\begin{align}
\label{first law matter}
    \mathrm{d}E_m=T\mathrm{d}S_m-P\mathrm{d}V_A,
\end{align}
where $E_m$, $S_m$ and $P$ are the energy, the entropy and the pressure of matter, respectively. It is important to emphasize that, under the assumption of thermal equilibrium, the gravitational and matter entropy contributions can be combined into a single expression. Therefore, the GFL for a cosmic model can be written as follows 
\begin{align}
    \label{f(A) GFL}
    \mathrm{d}E_T=T\mathrm{d}S_T+W_T \mathrm{d}V_A,
\end{align}
where the total energy $E_T$, the total entropy $S_T$, and the total work $W_T$ are given by 
\begin{align}
\label{f(A) total energy}
    \mathrm{d}E_T&=\mathrm{d}E_m+f'(A_A)\mathrm{d}E_g,\\
    \label{f(A) total entropy}
    S_T&=S_m+S_g,\\
    \label{f(A) total work}
    W_T&=-P+\tilde{W}f'(A_A).
\end{align}
This GFL is valid to any effective or alternative (quantum) cosmological model that can be written as in the standard cosmology taking as ansatz entropy as a function of the apparent area.
\subsection{Generalized Second Law}
\label{subsec: GSL f(A)}
The derivative of entropy can be expressed in terms of the Hubble parameter. Therefore, the second law for the gravitational part and for the matter part can be written as  
\begin{align}
    \label{f(A) second law g}
    \dot{S}_g&=2\pi R_A \dot{R}_Af'(A_A)=-2\pi \dfrac{H\left(\dot{H}-\dfrac{k}{a^2}\right)}{\left(H^2+\dfrac{k}{a^2}\right)^2}f'(A_A),\\
    \label{f(A) second law m}
    \dot{S}_m&=\frac{1}{T}\left[V_A\dot{\rho}+(\rho+P)\dot{V}_A\right]=-\dfrac{4\pi (P+\rho)H}{T\left(H^2+\dfrac{k}{a^2}\right)^{5/2}}\left[\dot{H}+H^2\right],\nonumber\\
    &=-\dfrac{16\pi(P+\rho)H(\dot{H}+H^2)}{\left\vert2H^2+\dot{H}+\dfrac{k}{a^2}\right\vert\left(H^2+\dfrac{k}{a^2}\right)^2}.
\end{align} 
Thus, using (\ref{f(A) total entropy}) the generalized second law is written as
\begin{align}
    \label{GSL k}
    \dot{S}_T&=\dot{S}_g+\dot{S}_m,
\end{align}
the validity region of the GSL comes when $\dot{S}_T\geq 0$. Using the SEC, the second law for the matter part (\ref{f(A) second law m}) is valid when $H^2+\dot{H}\leq 0$. On the other hand, $\dot{S}_m<0$ when $H^2+\dot{H}> 0$. The second law for the gravitational part (\ref{f(A) second law g}) holds ($\dot{S}_g\geq 0$) if $\dot{H}-\frac{k}{a^2}>0$ and $f'(A_A)<0$. Thus, GSL is valid if $\dot{S}_g>\dot{S}_m$. Also, the gravitational part obeys the second law if $\dot{H}-\frac{k}{a^2}<0$ and $f'(A_A)>0$. Here, GSL is valid if $H^2+\dot{H}< 0$ or if $\dot{S}_g>\dot{S}_m$ for $H^2+\dot{H}> 0$. 

The condition to get the sign of $\dot{H}-\frac{k}{a^2}$ is equivalent to showing that $\rho_\mathrm{eff}+P_\mathrm{eff}$ has the opposite sign. Furthermore, the condition to find the sign of $H^2+\dot{H}$ corresponds to proof that $\frac{\ddot{a}}{a}$ has the same sign or that $\rho_\mathrm{eff}+3P_\mathrm{eff}$ has the opposite sign because they are related by the effective Raychaudhuri equation (\ref{eff Raychaudhuri eq}). In contrast to standard cosmology, in effective models, the weak energy condition (WEC) is fulfilled because $\rho_\mathrm{eff}\geq 0$, but the null energy condition (NEC) may not be valid because $\rho_\mathrm{eff}+P_\mathrm{eff}$ can take any value, not only positive values. In addition, the strong energy condition (SEC), which would mean $\rho_\mathrm{eff}+3P_\mathrm{eff}\geq 0$ when $\rho_\mathrm{eff} +P_\mathrm{eff}\geq 0$ and $\rho_\mathrm{eff}\geq 0$, may not be valid. However, consistently with ordinary matter, we assume that the WEC, NEC, and SEC are valid for $\rho$ and $P$.

The GFL and GSL for effective cosmological models are obtained due to rewriting them as in standard cosmology \cite{Li-Thermo:2008tc,Zhang-Thermo:2021umq,Zhang-infla:2007bi}. The conditions of validity for the GSL depend on the effective model that we work with and the function of apparent area that we choose for the gravitational entropy. We will focus on effective models in LQC, in particular for the flat case with logarithmic corrections. However, this strategy can be used to compare and analyze alternative (quantum) cosmology models.

\section{Logarithmic corrections to entropy}
\label{section: log corrections eff}
In order to fix some function for the gravitational entropy, we consider the logarithmic correction to the entropy for an effective model; these corrections come from the counting of microstates of LQG black holes \cite{Meissner:2004ju,Ghosh:2004rq,Ghosh:2006ph,Agullo:2010zz-entropyBH,Engle:2009vc,Domagala:2004jt}. In LQG, gravitational entropy takes the following form  
\begin{align}
    \label{LQG entropy}
    S_g=\dfrac{ A_A}{4}+\tilde{\alpha}\ln \left(\dfrac{A_A}{4}\right)+\beta,
\end{align}
where $\tilde{\alpha}=\pi \alpha $, with $\alpha \sim \mathcal{O}(1)$ and $\beta$ being real constants from LQG corrections. 
\subsection{Generalized First Law.}
\label{subsec: GFL log correc}
As in Section \ref{section:Entropy in FLRW}, we can find the first law of thermodynamics for the gravitational part of effective models in cosmology using the LQG entropy (\ref{LQG entropy}) 
\begin{align}
\label{tilde first law}
    \mathrm{d}\tilde{E}_g=T\mathrm{d}S_g+\tilde{W}\mathrm{d}\tilde{V}_A-\dfrac{8\pi\alpha}{3}\tilde{W}\mathrm{d}R_A,
\end{align}
where $\tilde{W}$ is the effective work (\ref{eff work}), the gravitational energy, and the apparent volume with quantum corrections, $\tilde{E}_g$ and $\tilde{V}_A$, are defined as
\begin{align}
\label{tilde MS energy}
\tilde{E}_g&=\rho_\mathrm{eff}\tilde{V}_A,\\
\label{tilde apparent volume}
\tilde{V}_A&=V_A+\dfrac{4\pi\alpha}{3}R_A.
\end{align}
We obtain the correct limit when $\alpha\to 0$ and we recover the first law in \cite{Li-Thermo:2008tc,Zhang-Thermo:2021umq}, where $S_g\to A_A/4$, $\tilde{V}_A\to V_A$, and $\tilde{E}_g\to E_g$. Although the first law of LQC is read as in (\ref{tilde first law}), to include the matter part it is convenient to rewrite it as in (\ref{f(A) GFL})
\begin{align}
\label{D log First law}
    \mathrm{D}E_g=T\mathrm{d}S_g+\tilde{W}\mathrm{D}V_A,
\end{align}
where 
\begin{align}
    \mathrm{D}E_g=\left(1+\dfrac{4\pi\alpha}{A_A}\right)\mathrm{d}E_g,\nonumber\\
    \mathrm{D}V_A=\left(1+\dfrac{4\pi\alpha}{A_A}\right)\mathrm{d}V_A.
\end{align}
Hence, for $\alpha\to 0$ or for $A_A$ large, it is possible to get the standard first law of classical cosmology because of $S_g\to A_A/4$, $\mathrm{D}E_g\to\mathrm{d}E_g$, and $\mathrm{D}V_A\to\mathrm{d}V_A$.

The contribution of matter to analyze the generalized first law obeys the SEC (\ref{first law matter}). Due to the thermal equilibrium, the GFL for effective cosmological models can be written as follows
\begin{align}
    \label{GFL of LQC}
    \mathrm{d}E_T=T\mathrm{d}S_T+W_T \mathrm{d}V_A,
\end{align}
where the total energy $E_T$, the total entropy $S_T$ and the total work $W_T$ are given by 
\begin{align}
\label{f(A) total energy}
    \mathrm{d}E_T&=\mathrm{d}E_m+\left(1+\dfrac{4\pi\alpha}{A_A}\right)\mathrm{d}E_g,\\
    \label{total entropy}
    S_T&=S_m+S_g,\\
    \label{total work}
    W_T&=-P+\tilde{W}\left(1+\dfrac{4\pi\alpha}{A_A}\right).
\end{align}
The GFL of effective cosmological models using the logarithmic corrections can be written defining consistently an effective energy density and an effective pressure \cite{Li-Thermo:2008tc, Zhang-Thermo:2021umq}.

\subsection{Generalized Second Law.}
\label{subsec: GSL log correc}
The second law of thermodynamics with logarithmic corrections to entropy for the matter part is given in (\ref{f(A) second law m}), where the conditions of validity are the same as in Section \ref{subsec: GSL f(A)}. Nevertheless, the gravitational part reads as follows 
\begin{align}
    \label{second law g LQC}
    \dot{S}_g&=-2\pi \dfrac{H\left(\dot{H}-\dfrac{k}{a^2}\right)}{\left(H^2+\dfrac{k}{a^2}\right)^2}\left[1+\alpha\left(H^2+\dfrac{k}{a^2}\right)\right],
\end{align} 
we have $\dot{S}_g\geq 0$ when $\dot{H}-\frac{k}{a^2}>0$ and $1-\alpha\left(H^2+\frac{k}{a^2}\right)<0$ or when $\dot{H}-\frac{k}{a^2}<0$ and $1-\alpha\left(H^2+\frac{k}{a^2}\right)>0$. Therefore, using (\ref{f(A) second law m}) and (\ref{total entropy}) the GSL is written as
\begin{align}
    \label{GSL k}
    \dot{S}_T&=\dot{S}_g+\dot{S}_m.
\end{align}
The GSL with logarithmic corrections is valid when $\dot{H}+H^2\leq 0$ and $1-\alpha\left(H^2+\frac{k}{a^2}\right) >0$ or when $\dot{H}+H^2> 0$ and $\dot{S}_g>\dot{S}_m$ if the gravitational second law is fulfilled.

\section{Loop Quantum Cosmology.}
\label{sec: LQC}
In this Section, a brief review of the flat \cite{Ashtekar:2006wn,Ashtekar:2007em,Ashtekar:2003hd}, open \cite{Vandersloot:2006ws,Szulc:2007uk}, and closed \cite{Ashtekar:2006es,Corichi:2013usa,Corichi:2011pg,Szulc:2006ep,Gordon:2020gel,Li:2021fmu,Giesel:2022rxi} models in loop quantum cosmology is given. Particularly, the semiclassical description is revisited for those models where effective equations in LQC are written in the standard form of cosmology. Furthermore, we find a generalization for the modified Friedmann equations in LQC using a curvature parameter $k=0,\pm 1$. Despite writing the effective Friedmann equations as a unified description, there exist alternative models in LQC that include some ambiguities; for instance, the inclusion of the Lorentzian term in the Hamiltonian constraint. An alternative flat LQC model \cite{Yang:2009fp,Assanioussi:2018hee,Assanioussi:2019iye,Zhang:2021zfp,Gallegos:2024ajc} was rewritten in standard form in \cite{Zhang-Thermo:2021umq} for the GFL; the study of the GSL will remain as future work. This method could also be applied for the GFL and the GSL of the alternative $k=\pm 1$ LQC models \cite{Yang:2022aec,Gallegos:2026}. 

Loop Quantum Cosmology applies LQG techniques to a cosmological model, where a reduced-symmetry model by isotropic and homogeneity is transformed into Ashtekar-Barbero variables and then promoted to flux and holonomy variables to loop quantize finding a Hilbert space representation of this algebra. From the quantum description, an effective Hamiltonian constraint $C_{\mathrm{eff}}$ can be obtained using different methods \cite{Yang:2009fp,Assanioussi:2018hee,Assanioussi:2019iye,Ashtekar:2010gz,Qin:2012gaa,Huang:2011es,Ashtekar:2015iza,Corichi:2011sd}. The loop quantization shows that both quantum and effective descriptions remove the classical singularity through the quantum bounce.

The semiclassical description is expressed in terms of continuous variables, but includes quantum corrections arising from loop quantization. It is convenient to use this form of the cosmological equations because it allows a direct comparison between the classical and effective descriptions in cosmology.

\subsection{Flat LQC model}
The semiclassical description of the flat LQC model is obtained from the effective Hamiltonian constraint \cite{Ashtekar:2003hd,Ashtekar:2006wn,Ashtekar:2007em,Ashtekar:2010gz}
\begin{align}
  \label{flat effective constraint}
    C_\mathrm{eff}=-\frac{3}{8 \pi \gamma^2 \bar{\mu}^2} \sqrt{p} \sin ^2(\bar{\mu} c)+C_\phi,
    \end{align}
where $p_\phi$ is the momentum of the massless scalar field $\phi$, $C_\phi=\frac{1}{2}B(p)p_\phi^2$ and $B(p)$ are the eigenvalues of $\widehat{p^{-3/2}}$, and $\gamma\approx 0.23$ is the Barbero-Immirzi parameter. This value is fixed using the entropy calculus for black holes in LQC \cite{Meissner:2004ju,Ghosh:2004rq,Rovelli:1996dv,Ashtekar:1997yu,Agullo:2010zz-entropyBH,Ghosh:2006ph}. Moreover, $c,p$ are the coordinates of the symmetry-reduced gravitational phase space, which obey Poisson brackets 
\begin{align}
\label{Poisson brackets c,p}
    \{c, p\}=\frac{8 \pi G \gamma}{3},
\end{align}
and $\bar{\mu}^2 p=\Delta$, where $\bar{\mu}$ is a nontrivial function on the phase space, and $\Delta=2\sqrt{3}\pi \gamma \ell_{p}^2$ is the non-zero minimum eigenvalue of the area operator. The equations of motion are obtained using the Hamiltonian equations in (\ref{flat effective constraint})
\begin{align}
    \dot{v}&=\{v,C_\mathrm{eff}\}=-\dfrac{8\pi\gamma G}{3}\dfrac{\partial C_\mathrm{eff}}{\partial c},\\
    \dot{\phi}&=\{\phi,C_\mathrm{eff} \},
\end{align}
where $v$ is the flux variable, which is proportional to the eigenvalues  of the volume operator $\widehat{V}$. Hence, the modified Friedmann equation is obtained 
\begin{equation}
\label{flat modified Friedmann eq}
    H^2=\left(\dfrac{\dot{v}}{3v}\right)^2=\dfrac{8\pi G}{3}\rho\left(1-\dfrac{\rho}{\rho_c}\right),
\end{equation}
where $H=\frac{\dot{a}}{a}$ is the Hubble parameter, $\rho$ and $\rho_c$ are the energy density and critical energy density in flat curvature, respectively, which are defined as 
\begin{align}
\label{rho}
\rho&=\dfrac{p_\phi}{2p^3},\\
    \label{critical rho}
    \rho_c&=\dfrac{3}{8\pi G\gamma^2 \Delta},
\end{align}
when the energy density is equal to the critical one, a quantum bounce occurs, and when $\rho<<\rho_c$ quantum effects are negligible. Thus, the classical description is recovered.
In addition, the pressure of matter is given by 
\begin{align}
    P=-\dfrac{\partial C_\phi}{\partial V},
\end{align}
we can rewrite (\ref{flat modified Friedmann eq}) as the standard form in cosmology
\begin{equation}
    \label{eff flat modified Friedmann eq}
    H^2=\dfrac{8\pi G}{3}\rho_\mathrm{eff},
\end{equation}
where the effective energy density is defined as 
\begin{align}
\label{eff flat energy density}
    \rho_\mathrm{eff}=\rho\left(1-\dfrac{\rho}{\rho_c}\right),
\end{align}
using the continuity equation (\ref{FLRW continuity eq}) and the modified Friedmann equation (\ref{eff flat modified Friedmann eq}), it is possible to obtain the effective Raychaudhuri (\ref{eff Raychaudhuri eq}) and effective conservation (\ref{eff continuity eq}) equations, where the effective pressure is defined as  
\begin{align}
\label{eff k=-1 pressure}
    P_{\mathrm{eff}}=P\left(1-\frac{2 \rho}{\rho_c}\right)-\frac{\rho^2}{\rho_c}.
\end{align}
Furthermore, we can analyze (\ref{eff dot H}) for this flat LQC model 
\begin{align}
\label{flat acc modif hubble}
    \dot{H}=-4\pi (P+\rho)\left(1-\dfrac{2\rho}{\rho_c}\right).
\end{align}
We note that in (\ref{flat acc modif hubble}) $\dot{H}<0$ for the region $0<\rho<\frac{\rho_c}{2}$ and $\dot{H}>0$ for $\frac{\rho_c}{2}<\rho<\rho_c$. Here, the weak energy condition (WEC) is valid; it means $\rho\geq 0$ and the null energy condition (NEC) is obeyed, where $\rho+P\geq 0$.

\subsection{Open LQC model}
\label{sub: open LQC model}
The case when $k=-1$ is analyzed using the modified Friedmann equations \cite{Vandersloot:2006ws,Szulc:2007uk} where the spatial curvature plays an important role. Here, 
\begin{align}
    \label{k=-1 modified Friedmann eq}
    H^2=\left(\frac{8\pi G}{3} \rho+\frac{1}{a^2}\right)\left(1-\frac{\rho}{\rho_c}-\frac{\gamma^2 \Delta}{a^2} \right),
\end{align}
comparing with the standard form of the Friedmann equation (\ref{eff friedmann eq}), the quantum contribution is written between the second parenthesis in (\ref{k=-1 modified Friedmann eq}). Furthermore, the contribution of spatial curvature is encoded in $\mathcal{O}(a^{-2})$ terms; if we do not consider these terms, we have a flat modified Friedmann equation (\ref{flat modified Friedmann eq}). 

We can rewrite (\ref{k=-1 modified Friedmann eq}) in the standard way as follows 
\begin{align}
    \label{eff k=-1 Friedamann eq}
    H^2=\dfrac{8\pi G}{3}\rho_\mathrm{eff}+\dfrac{1}{a^2},
\end{align}
where the effective energy density is given by
\begin{align}
\label{eff k=-1 energy density}
    \rho_{\text {eff }}=\rho\left(1-\frac{\rho}{\rho_c}-\frac{\gamma^2 \Delta}{a^2}\right)-\frac{3}{8 \pi G a^2}\left(\frac{\gamma^2 \Delta}{a^2}+\frac{\rho}{\rho_c}\right).
\end{align}

To rewrite the effective Raychaudhuri and effective continuity equations as in (\ref{eff Raychaudhuri eq}) and (\ref{eff continuity eq}), we define the effective pressure as 
\begin{align}
    \label{eff k=-1 preassure}
    P_{\text {eff }}&=P\left(1-\frac{2 \rho}{\rho_c}-\frac{\gamma^2 \Delta}{a^2}\right)-\rho\left(\frac{\rho}{\rho_c}+\frac{2 \gamma^2 \Delta}{3 a^2}\right)\nonumber\\
    &-\frac{1}{8 \pi G a^2}\left(\frac{\gamma^2 \Delta}{a^2}+\frac{2\rho}{\rho_c}+\dfrac{3P}{\rho_c}\right),
\end{align}
with the definition of $\rho_\mathrm{eff}$ and $P_\mathrm{eff}$, we can rewrite effective equations for $k=-1$ in the form of eqs. (\ref{eff k=-1 Friedamann eq}), (\ref{eff Raychaudhuri eq}), (\ref{eff continuity eq}), and (\ref{eff dot H}).

Furthermore, the quantum bounce occurs when $H=0$, that is, when the energy density reaches its maximum value 
\begin{align}
    \label{k=-1 maximum rho}
    \rho_*=\rho_c\left(1-\dfrac{\gamma^2\Delta}{a^2_{min}}\right),
\end{align}
where $\rho_c$ is the critical energy density in the flat model (\ref{critical rho}) and $a_{min}$ is the minimum value of the scale factor when the quantum bounce occurs. For this model, the expression (\ref{eff dot H}) is read as 
\begin{align}
\label{dot H k=-1}
\dot{H}+\frac{1}{a^2}=\left(-4 \pi G(\rho+P)-\frac{1}{a^2}\right)\left(1-\frac{2 \rho}{\rho_c}-\frac{\gamma^2 \Delta}{a^2}\right)+\frac{1}{a^2}\left(1+\frac{\rho+3 P}{\rho_c}+\frac{\gamma^2 \Delta}{a^2}+\frac{8 \pi G \rho \gamma^2 \Delta}{3}\right),
\end{align}
when the WEC and the strong energy condition (SEC) are fulfilled, that is, $\rho+3P\geq 0$ and $\rho+P\geq 0$. We have that  $\dot{H}+\frac{1}{a^2}\geq 0$ when $\rho \geq\frac{\rho_*}{2}$. On the other hand, $\dot{H}+\frac{1}{a^2}<0$ when $\rho<\frac{\rho_*}{2}$. These conditions, when the quantity $\dot{H}+\frac{1}{a^2}$ changes sign, are equivalent to the flat LQC case.

\subsection{Closed LQC model}
\label{sub: closed LQC model}

The closed case in LQC has been extensively studied. However, in this section, we focus on the effective description of the $k=+1$ model \cite{Ashtekar:2006es,Corichi:2011pg,Corichi:2013usa,Li:2021fmu,Giesel:2022rxi,Gordon:2020gel,Szulc:2006ep}. The effective Hamiltonian constraint including the matter part $C_\phi$ is given by 
\begin{align}
\label{eff constraint k=+1}
    C_{\mathrm{eff}}=-\frac{3 v}{8 \pi G \gamma^2 \lambda^2}\left[\sin ^2(\lambda b-D)-\sin ^2 D+\left(1+\gamma^2\right) D^2\right]+C_\phi,
\end{align}
where $D=\lambda\left(\frac{2 \pi^2}{v}\right)^{1 / 3}$, and $\lambda^2=\Delta$. The modified Friedmann equation can be obtained from the Hamilton equations of (\ref{eff constraint k=+1}), such that
\begin{align}
\label{eff friedmann k=+1 a}
    H^2=\frac{\dot{v}^2}{9 v^2}=\frac{8 \pi G}{3}\left(\rho-\rho_{\min }\right)\left(1-\frac{\rho-\rho_{\min }}{\rho_c}\right),
\end{align}
where the minimum energy density is defined as 
\begin{align}
    \rho_{\min }=\rho_c\left[\left(1+\gamma^2\right) D^2-\sin ^2 D\right] .
\end{align}
during all evolution $\rho_{\min } \leq \rho \leq \rho_{\max }$, where $\rho_{\max }=\rho_{\min }+\rho_c$. In contrast to the $k=0,-1$ LQC models, the closed model contains more than one bounce, when $\rho=\rho_\mathrm{min}$ and when $\rho=\rho_\mathrm{max}$. It means that, depending on the initial conditions, the Universe has either a bounce or a collapse when the energy density takes a minimum or maximum value. \\
Furthermore, in the limit when $|v|>>1$ and $\rho_\mathrm{min}\approx\rho_{\text {crit }} \gamma^2 D^2=\frac{3}{8 \pi G a^2}$, the equation (\ref{eff friedmann k=+1 a}) takes the following expression 
\begin{align}
\label{eff Friedmann k=+1 b}
    H^2&=\frac{8 \pi G}{3}\left(\rho-\frac{\rho^2}{\rho_c}+2 \rho \gamma^2 D^2-\rho_c \gamma^2 D^2-\rho_c \gamma^4 D^4\right)\nonumber\\
    &=\dfrac{8\pi G}{3}\rho_\mathrm{eff}-\dfrac{1}{a^2},
\end{align}
where 
\begin{align}
\label{eff k=+1 energy density}
    \rho_{\text {eff }}=\rho\left(1-\frac{\rho}{\rho_c}+\frac{\gamma^2 \Delta}{a^2}\right)+\frac{3}{8 \pi G a^2}\left(\frac{\rho}{\rho_c}-\frac{\gamma^2 \Delta}{a^2}\right).
\end{align}
Additionally, when $H=0$ the critical $k=+1$ energy density is defined  
\begin{align}
    \label{k=+1 rho_*}
    \rho_*=\rho_c\left(1-\dfrac{\gamma^2\Delta}{a^2_{min}}\right).
\end{align}
We define a consistent effective pressure with which it is possible to rewrite the effective Raychaudhuri (\ref{eff Raychaudhuri eq}) and effective continuity equations (\ref{eff continuity eq}) as follows
\begin{align}
    \label{eff k=+1 preassure}
    P_{\text {eff }}&=P\left(1-\frac{2 \rho}{\rho_c}+\frac{\gamma^2 \Delta}{a^2}\right)-\rho\left(\frac{\rho}{\rho_c}-\frac{2 \gamma^2 \Delta}{3 a^2}\right)\nonumber\\
    &-\frac{1}{8 \pi G a^2}\left(-\frac{\gamma^2 \Delta}{a^2}+\frac{2\rho}{\rho_c}+\dfrac{3P}{\rho_c}\right).
\end{align}
For the effective $k=+1$ LQC model, the expression (\ref{eff dot H}) is given by 

\begin{align}
\label{dot H k=+1}
\dot{H}-\frac{1}{a^2}=\left(-4 \pi G(\rho+P)+\frac{1}{a^2}\right)\left(1-\frac{2 \rho}{\rho_c}+\frac{\gamma^2 \Delta}{a^2}\right)+\frac{1}{a^2}\left(1+\frac{\rho+3 P}{\rho_c}-\frac{\gamma^2 \Delta}{a^2}+\frac{8 \pi G \rho \gamma^2 \Delta}{3}\right),
\end{align} 
if the WEC and the SEC are applied to this case, then we can determine that $\dot{H}-\frac{1}{a^2}> 0$ either if $4\pi G(\rho+P)>\frac{1}{a^2}$ and $\rho<\frac{\rho_*}{2}$ or if $4\pi G(\rho+P)<\frac{1}{a^2}$ and $\rho>\frac{\rho_*}{2}$. Similarly, for $\dot{H}-\frac{1}{a^2}< 0$. This is possible when 
\begin{align}
    \left(-4 \pi G(\rho+P)+\frac{1}{a^2}\right)\left(1-\frac{2 \rho}{\rho_c}+\frac{\gamma^2 \Delta}{a^2}\right)<\frac{1}{a^2}\left(1+\frac{\rho+3 P}{\rho_c}-\frac{\gamma^2 \Delta}{a^2}+\frac{8 \pi G \rho \gamma^2 \Delta}{3}\right),
\end{align}
that is when $\left(-4 \pi G(\rho+P)+\frac{1}{a^2}\right)\left(1-\frac{2 \rho}{\rho_c}+\frac{\gamma^2 \Delta}{a^2}\right)\geq 0$ or $\left(-4 \pi G(\rho+P)+\frac{1}{a^2}\right)\left(1-\frac{2 \rho}{\rho_c}+\frac{\gamma^2 \Delta}{a^2}\right)<0$. The first case is equivalent to if $4\pi G(\rho+P)>\frac{1}{a^2}$ and $\rho<\frac{\rho_*}{2}$ or if $4\pi G(\rho+P)<\frac{1}{a^2}$ and $\rho>\frac{\rho_*}{2}$. The second case occurs if $4\pi G(\rho+P)<\frac{1}{a^2}$ and $\rho<\frac{\rho_*}{2}$ or if $4\pi G(\rho+P)>\frac{1}{a^2}$ and $\rho>\frac{\rho_*}{2}$.
\subsection{Unified effective description in LQC}
\label{sub: unified LQC}
The effective models in LQC have been revisited in previous subsections, where the modified Friedmann equations can have a unified description comparing (\ref{flat modified Friedmann eq}), (\ref{k=-1 modified Friedmann eq}) and (\ref{eff Friedmann k=+1 b}) as follows 
\begin{align}
\label{unified Friedmann equation}
    H^2=\left(\frac{8 \pi G}{3} \rho-\frac{k}{a^2}\right)\left(1-\frac{\rho}{\rho_c}+\frac{k\gamma^2 \Delta}{a^2}\right).
\end{align}
Despite the fact that the topology and the deduction of each case are fundamentally different, this equation generalizes every case with curvature using a parameter $k$ , and it is possible to rewrite it in the standard cosmological way (\ref{eff friedmann eq}) if the effective energy density and the effective pressure are defined as 
\begin{align}
\label{eff unified energy density}
    \rho_{\text {eff }}&=\rho\left(1-\frac{\rho}{\rho_c}+\frac{k\gamma^2 \Delta}{a^2}\right)+\frac{3}{8 \pi G a^2}\left(\frac{\rho}{\rho_c}-\frac{k\gamma^2 \Delta}{a^2}\right),\\
      \label{eff unified preassure}
    P_{\text {eff }}&=P\left(1-\frac{2 \rho}{\rho_c}+\frac{k\gamma^2 \Delta}{a^2}\right)-\rho\left(\frac{\rho}{\rho_c}-\frac{2 k\gamma^2 \Delta}{3 a^2}\right)\nonumber\\
    &-\frac{1}{8 \pi G a^2}\left(-\frac{k\gamma^2 \Delta}{a^2}+\frac{2\rho}{\rho_c}+\dfrac{3P}{\rho_c}\right), 
\end{align}
these definitions are consistent with writing the standard cosmic equations, and the modified Friedmann equations reduce to the classical Friedmann equations in cosmology for each case where the appropriate limit is taken; this means $ \rho \ll 1 $ and $ a \gg 1$. On the other hand, the critical energy density $\rho_*$ is found when $H=0$
\begin{align}
    \label{unified rho_*}
    \rho_*=\rho_c\left(1-\dfrac{k\gamma^2\Delta}{a^2_{min}}\right).
\end{align}
Note that for the flat case $\rho_*$ is exactly $\rho_c$, and the critical energy density reduces in a fashion similar to the other cases. Additionally, an important quantity to find the validity regions for the GSL in LQC is given by 
\begin{align}
\label{dot H unified}
\dot{H}=\left(-4 \pi G(\rho+P)+\frac{1}{a^2}\right)\left(1-\frac{2 \rho}{\rho_c}+\frac{k\gamma^2 \Delta}{a^2}\right)+\frac{k}{a^2}\left(\frac{\rho+3 P}{\rho_c}-\frac{k\gamma^2 \Delta}{a^2}+\frac{8 \pi G \rho \gamma^2 \Delta}{3}\right),
\end{align} 
implementing the WEC and the SEC, the change of sign for $\dot{H}-\frac{k}{a^2}$ has been analyzed for each case previously. Note that for the flat case, the SEC was not used; it was sufficient to use the WEC. \\
The WEC for the effective equations is valid because $H^2+\frac{k}{a^2}\geq 0$, then $\rho_\mathrm{eff}\geq 0$. On the other hand, $\dot{H}-\frac{k}{a^2}$ can take any value. Therefore, if $\dot{H}-\frac{k}{a^2}\geq 0$ implies $H^2 +\dot{H}\geq 0$, and if $\dot{H}-\frac{k}{a^2}<0$, there exist two possibilities: either $H^2+\dot{H}\geq 0$ or $H^2+\dot{H}<0$. Thus, using WEC, we note that if $\dot{H}-\frac{k}{a^2}\geq 0$ implies $2H^2+\dot{H}+\frac{k}{a^2}\geq H^2+\frac{k}{a^2}\geq 0$. In addition, if $\dot{H}-\frac{k}{a^2}<0$ and $H^2+\dot{H}\geq 0$, then $2H^2+\dot{H}+\frac{k}{a^2}\geq 0$; but if $\dot{H}-\frac{k}{a^2}<0$ and $H^2+\dot{H}< 0$, then $2H^2+\dot{H}+\frac{k}{a^2}< H^2+\frac{k}{a^2}$. Namely, $2H^2+\dot{H}+\frac{k}{a^2}$ can take negative, zero, or positive values. This analysis will be of interest for calculating the regions where the GSL is valid.

\section{Generalized First Law in LQC.}
\label{subsec: GFL in LQC}
The GFL in LQC is described by the eq. (\ref{GFL of LQC}), where the effective energy density (\ref{eff unified energy density}) and the effective pressure (\ref{eff unified preassure}) are considered. The apparent horizon is used as in the general form (\ref{apparent radius}). As a particular case, the GFL is considered in the flat LQC, which has the form in (\ref{GFL of LQC}) with $R_A=1/H$. Logarithmic corrections are encoded in the tilde quantities, which are relevant to small area scales. In this case, the effective work (\ref{eff work}) can be expressed as: 
\begin{align}
    \tilde{W}=W+\dfrac{\rho}{\rho_c}P.
\end{align}
Hence, the first law of the gravitational part can be rewritten as 
\begin{align}
\label{LQC tilde first law}
    \mathrm{d}\tilde{E}_g=T\mathrm{d}S_g+\left(W+\dfrac{\rho}{\rho_c}P\right)\mathrm{d}\left(\tilde{V}_A-\dfrac{8\pi\alpha}{3}R_A\right).
\end{align}
The first law of the gravitational part without logarithmic correction ($\alpha=0$) \cite{Li-Thermo:2008tc} can be obtained because of $S_g\to A_A/4$, $\tilde{V}_A\to V_A$, and $\tilde{E}_g\to E_g$. For the case when $\rho<<\rho_c$, we recover the standard expression in cosmology, and quantum effects are negligible.

\section{Generalized Second in LQC.}
\label{subsec: GSL in LQC}
One of the main results in this article is the study of the GSL of thermodynamics for LQC, which is given by (\ref{GSL k}), where the Hubble parameter depends on the effective Friedmann equation of each model. In this section, the $k=0,\pm 1$ LQC models are discussed, and the validation conditions are explored using the logarithmic corrections due to the black hole entropy calculation from LQG. 
\subsection{GSL for a flat LQC model}
Validation conditions for the GSL of a flat LQC model with logarithmic corrections are analyzed for every possible value of $\tilde{\alpha}$. The second law for gravity and matter in this case is written as 
\begin{align}
\label{dot Sg}
    \dot{S}_g&=-2\pi \dfrac{\dot{H}}{H^3}(1+\alpha H^2),\\
\label{dot Sm}
    \dot{S}_m&=-\dfrac{4\pi (P+\rho)}{TH^4}\left(H^2+\dot{H}\right),\nonumber\\
    &=4\pi \dfrac{\dot{H}}{H^3}\dfrac{\dot{H}+H^2}{\left\vert 2H^2+\dot{H}\right\vert}\left(\dfrac{\rho_c}{\rho_c-2\rho}\right),
\end{align}
the GSL for LQC is given by 
\begin{align}
\label{LQC GSL}
    \dot{S}_T&=\dot{S}_g+\dot{S}_m,\nonumber\\
    &=-2\pi \dfrac{\dot{H}}{H^3}\left[(1+\alpha H^2)-\dfrac{\dot{H}+H^2}{\left\vert 2H^2+\dot{H}\right\vert}\left(\dfrac{2\rho_c}{\rho_c-2\rho}\right)\right].
\end{align}

The GSL is valid when $\dot{S}_T\geq 0$, we can express the pressure in terms of the energy density as $P=w\rho$ to get the following state equation, where the validity regions are equivalent to solving the following expression
\begin{align}
    \label{GSL state eq}
    \dfrac{\dot{H}}{H^3}\left[-\left(1+\dfrac{8\pi\alpha}{3}\rho_c x(1-x)\right)
    +\left(\dfrac{2x-3(w+\frac{1}{3})(1-2x)}{\left\vert 4x-3(w-\frac{1}{3})(1-2x)\right\vert}\right)\left(\dfrac{2}{1-2x}\right)\right]\geq 0,
\end{align}
where $x=\frac{\rho}{\rho_c}$, and the weak energy condition is valid when $\omega>-1$. Additionally, the second law for the matter part holds when $-\frac{2}{3}<\omega$ at $\rho_c/2<\rho<\rho_c$, that is $\dot{H}>0$ and at $0<\rho<\rho_c/2$, namely $\dot{H}<0$, we have $-\frac{2}{3}<\omega$ for $H^2+\dot{H}<0$ and $-1<\omega<-\frac{2}{3}$ for $H^2+\dot{H}>0$.

We study every possible value that the factor in the logarithmic correction $\tilde{\alpha}=\pi\alpha$ can have.

\textbf{Case $\alpha=0$}. The first case considers when there is no logarithmic correction. Conditions of validity of (\ref{LQC GSL}) taking into account (\ref{flat acc modif hubble}) at $0<\rho<\rho_c/2$ are given by
\begin{itemize}
  \item $\dot{S}_T\geq 0$ when $\dot{S}_g\geq \dot{S}_m$ for $H^2+\dot{H}>0$ or equivalently using (\ref{GSL state eq}) we have
  \begin{align}
      \left(\frac{2 x-3\left(w+\frac{1}{3}\right)(1-2 x)}{\vert 4 x-3\left(w-\frac{1}{3}\right)(1-2 x)\vert}\right)\left(\frac{2}{1-2 x}\right) \leq 1.
  \end{align}
  \item $\dot{S}_T\geq 0$ for $H^2+\dot{H}<0$ because $\dot{S}_g\geq 0$ and $\dot{S}_m\geq 0$.
\end{itemize}
On the other hand, at $\rho_c/2<\rho<\rho_c$ we have $\dot{S}_T<0$ because $\dot{S}_g< 0$ and $\dot{S}_m< 0$.

\textbf{Case $\alpha>0$}. The positive logarithmic correction \cite{Hod:2004di} implies $1+\alpha H^2>0$. Thus, at $0<\rho<\rho_c/2$
\begin{itemize}
  \item $\dot{S}_T\geq 0$ when $\dot{S}_g\geq \dot{S}_m$ for $H^2+\dot{H}>0$, which corresponds to write the state equation as 
  \begin{align}
      \left(\frac{2 x-3\left(w+\frac{1}{3}\right)(1-2 x)}{\vert 4 x-3\left(w-\frac{1}{3}\right)(1-2 x)\vert }\right)\left(\frac{2}{1-2 x} \right) \leq 1+\alpha H^2.
  \end{align}
  \item $\dot{S}_T\geq 0$ for $H^2+\dot{H}<0$, because $\dot{S}_g\geq 0$ and $\dot{S}_m\geq 0$.
\end{itemize}
Similarly as in the case $\alpha=0$, at $\rho_c/2<\rho<\rho_c$ we find $\dot{S}_T< 0$.

\textbf{Case $\alpha_0=-\frac{3}{2\pi\rho_c}$}. The transition point $\rho_0=\frac{\rho_c}{2}$ determines the change of sign of $\dot{H}$ and it is possible to find $\alpha_0$ such that $1+\alpha_0 H^2(\rho_0)=0$, where $\alpha_0=-\frac{3}{2\pi G\rho_c}\approx -1.16$ \cite{Sadjadi:2012wg}. Note that the factor in the logarithmic correction is given by $\tilde{\alpha}_0=\pi\alpha_0\approx -3.64$. Additionally, the maximum value of $H^2$ is when $H^2(\rho_0)$, then we have $1+\alpha_0 H^2(\rho)>0$ for $\rho\neq \rho_0$. Hence, validity regions of the GSL at $0<\rho<\rho_0$
\begin{itemize}
  \item $\dot{S}_T\geq 0$ when $\dot{S}_g\geq \dot{S}_m$ for $H^2+\dot{H}>0$, equivalently when the state equation obeys the following expression
  \begin{align}
      \left(\frac{2 x-3\left(w+\frac{1}{3}\right)(1-2 x)}{\vert 4 x-3\left(w-\frac{1}{3}\right)(1-2 x)\vert }\right)\left(\frac{2}{1-2 x}\right) \leq 1-4x(1-x).
  \end{align}
  \item $\dot{S}_T\geq 0$ for $H^2+\dot{H}<0$, because $\dot{S}_g\geq 0$ and $\dot{S}_m\geq 0$.
\end{itemize}
Since, $\dot{S}_m<0$ and $\dot{S}_g<0$ at $\rho_0<\rho<\rho_c$, then the GSL is not valid. 

\textbf{Case $\tilde{\alpha}_0<\tilde{\alpha}<0$}. Logarithmic corrections due to the counting of the microstates of BH in LQG give as values $\tilde{\alpha}=-\frac{1}{2}$ \cite{Ashtekar:1997yu,Domagala:2004jt,Ghosh:2004rq,Agullo:2010zz-entropyBH,Ghosh:2006ph} and $\tilde{\alpha}=-\frac{3}{2}$ \cite{Engle:2009vc}. These values $\tilde{\alpha}_0<\tilde{\alpha}<0$. Assuming there exists a $\rho_1$ such that $\alpha_1=-\frac{1}{H^2(\rho_1)}$, it implies that $H^2(\rho_0)<H^2(\rho_1)$; it is not possible, because the maximum value of $H^2$ is at $\rho_0$. Therefore, the analysis of the GSL validation regions corresponds to the case $\alpha=\alpha_0$.

\textbf{Case $\tilde{\alpha}<\tilde{\alpha}_0<0$.} Examples to this case have not yet been found using the counting of microstates for BH in LQG. We assume $\alpha_2$ such that $\alpha_2=-\frac{1}{H^2(\rho_2)}$, then $H^2(\rho_2)<H^2(\rho_0)$ for $\alpha_2<\alpha_0<0$ . We have two possibilities, when $H^2(\rho)<H^2(\rho_2)<H^2(\rho_0)$ and when $H^2(\rho_2)<H^2(\rho)<H^2(\rho_0)$. 

At $0<\rho<\rho_0$,
\begin{itemize}
  \item We consider $H^2(\rho)<H^2(\rho_2)<H^2(\rho_0)$, this implies $1+\alpha_2 H^2>0$ and $\dot{S}_g<0$. We have 
  $\dot{S}_T\geq 0$ when $\dot{S}_g\geq \dot{S}_m$ for $H^2+\dot{H}>0$, equivalently when the following expression obeys
  \begin{align}
      \left(\frac{2 x-3\left(w+\frac{1}{3}\right)(1-2 x)}{\vert 4 x-3\left(w-\frac{1}{3}\right)(1-2 x)\vert }\right)\left(\frac{2}{1-2 x}\right) \leq 1+\frac{8\pi\alpha_2\rho_c}{3}x(1-x).
  \end{align}
  In addition, due to $\dot{S}_g\geq 0$ and $\dot{S}_m\geq 0$ for $H^2+\dot{H}<0$, we have $\dot{S}_T\geq 0$.
  \item For $H^2(\rho_2)<H^2(\rho)<H^2(\rho_0)$, it corresponds to $0>1+\alpha_2 H^2$ and $\dot{S}_g<0$. We have $\dot{S}_T\geq 0$ only when $\dot{S}_m\geq \dot{S}_g$; it is equivalent to proof that 
    \begin{align}
      \left(\frac{2 x-3\left(w+\frac{1}{3}\right)(1-2 x)}{\vert 4 x-3\left(w-\frac{1}{3}\right)(1-2 x)\vert }\right)\left(\frac{2}{1-2 x}\right) \leq 1+\frac{8\pi\alpha_2\rho_c}{3}x(1-x).
  \end{align}
Since $\dot{S}_m<0$, then $\dot{S}_T< 0$ for $H^2+\dot{H}>0$.
\end{itemize}
At $\rho_0<\rho<\rho_c$,
\begin{itemize}
    \item For $H^2(\rho_2)<H^2(\rho)<H^2(\rho_0)$, we have $\dot{S}_g>0$, and therefore $\dot{S}_T\geq 0$ only if $\dot{S}_g\geq\dot{S}_m$, or equivalently, the state equation takes the following form
    \begin{align}
      \left(\frac{2 x-3\left(w+\frac{1}{3}\right)(1-2 x)}{\vert 4 x-3\left(w-\frac{1}{3}\right)(1-2 x)\vert }\right)\left(\frac{2}{1-2 x}\right) \geq 1+\frac{8\pi\alpha_2\rho_c}{3}x(1-x).
  \end{align}
  \item For $H^2(\rho)<H^2(\rho_2)<H^2(\rho_0)$, then $1+\alpha_2 H^2>0$. Therefore, $\dot{S}_T< 0$ because $\dot{S}_g< 0$ and $\dot{S}_m<0$.
\end{itemize}

We consider every possible factor of the logarithmic correction $\tilde{\alpha}$ to entropy. Although there are conditions under which the GSL is valid, even when the second law for the gravitational part or the matter part fails. Note that the GSL is not valid just after the quantum bounce ($\rho_0<\rho<\rho_c$). Only in the case when $\tilde{\alpha}<\tilde{\alpha}_0<0$ does and $H^2(\rho_2)<H^2(\rho)<H^2(\rho_0)$, the GSL becomes valid just after the quantum bounce.\\
Due to the form of the matter entropy part for the flat model, the validity regions of the GSL consider only the WEC. It is not necessary to include NEC or SEC. 

\subsection{GSL for an open LQC model}
\label{Subsec: GSL k=-1} 
Using the effective equations for the open LQC model, which are shown in Section \ref{Subsec: GSL k=-1}, we can write the GFL as in (\ref{GFL of LQC}) for $k=-1$, where the apparent radius takes the following form $R_A=1/\sqrt{H^2-\frac{1}{a^2}}$
 
The second laws for the gravitational part and the matter part are given by 
\begin{align}
    \label{k=-1 second law g LQC}
    \dot{S}_g&=-2\pi \dfrac{H\left(\dot{H}+\dfrac{1}{a^2}\right)}{\left(H^2-\dfrac{1}{a^2}\right)^2}\left[1+\alpha\left(H^2-\dfrac{1}{a^2}\right)\right],\\
        \label{k=-1 second law m}
    \dot{S}_m&=-\dfrac{4\pi (P+\rho)H}{T\left(H^2-\dfrac{1}{a^2}\right)^{5/2}}\left[\dot{H}+H^2\right]\nonumber\\
    &=-\dfrac{16\pi(P+\rho)H(\dot{H}+H^2)}{\left\vert2H^2+\dot{H}-\dfrac{1}{a^2}\right\vert\left(H^2-\dfrac{1}{a^2}\right)^2}.
\end{align} 
Validity regions of the GSL ($\dot{S}_T\geq 0$) for the hyperbolic case in LQC are determined by the signs of $ \dot{H}+\frac{1}{a^2}$ and $\dot{H}+H^2$ for different values of the logarithmic correction $\alpha$, or equivalently, to find the signs of $\rho_\mathrm{eff}+P_\mathrm{eff}$ and $\rho_\mathrm{eff}+3P_\mathrm{eff}$ in different regions of $\rho$.\\
The WEC and SEC are implemented in this analysis. We find different cases using the value of $\alpha$ and study the validity regions in them.\\

\textbf{Case $\alpha=0$}. This case reproduces the results for the usual gravitational entropy definition in cosmology. Close to the quantum bounce, $\rho>\frac{\rho_*}{2}$, which is for $\dot{H}+\frac{1}{a^2}> 0$, implies that $\dot{S}_g<0$ and $\dot{S}_m<0$. Therefore, the GSL is violated.\\
Now, for $\dot{H}+\frac{1}{a^2}< 0$, when $\rho<\frac{\rho_*}{2}$, namely far away from the quantum bounce, we have $\dot{S}_g>0$. Here, we have two options, when $\dot{H}+H^2>0$, then $\dot{S}_m<0$. Thus, the GSL is only valid if $\dot{S}_g\geq \dot{S}_m$. The other option is given when $\dot{H}+\frac{1}{a^2}< 0$ and $\dot{H}+H^2<0$; this implies that $\dot{S}_T\geq 0$.\\
In this case, the GSL is violated close to the quantum bounce, and it is valid far away; this result is similar to the flat model in LQC.\\

\textbf{Case $1+\alpha(H^2-\frac{1}{a^2})\geq 0$}. This case includes the cases for $\alpha\geq 0$. Note that for $\dot{H}+\frac{1}{a^2}> 0$, this means that when $\rho\geq\rho_*/2$, we have $\dot{S}_T<0$.\\
When we have $\dot{H}+\frac{1}{a^2}< 0$, which is far away from the quantum bounce, the gravitational part is given by $\dot{S}_g> 0$. On the other hand, the matter part depends on two options. Firstly, when $\dot{H}+H^2>0$, this implies that $\dot{S}_m<0$, and the GSL is valid only if $\dot{S}_g\geq\dot{S}_m$. Secondly, when $\dot{H}+H^2<0$, we find that the GSL is fulfilled, namely $\dot{S}_T\geq 0$.\\
These results reproduce the previous analysis similarly when $\alpha=0$.\\

\textbf{Case $1+\alpha(H^2-\frac{1}{a^2})< 0$}. In this case, when we assume that $\dot{H}+\frac{1}{a^2}\geq 0$, it implies that $\dot{S}_g>0$ and $\dot{S}_m<0$. Hence, the GSL is valid only if $\dot{S}_g\geq\dot{S}_m$.\\
In contrast, when $\dot{H}+\frac{1}{a^2}< 0$, we find that $\dot{S}_g<0$. For the matter part, we study two possibilities: when $\dot{H}+H^2>0$, then $\dot{S}_m<0$. Thus, the GSL is not valid. Now, when  $\dot{H}+H^2<0$, we have $\dot{S}_m>0$. Therefore, $\dot{S}_T\geq 0$, when $\dot{S}_m\geq \dot{S}_g$. \\
For this case, we note that the violation of the GSL is not close to the quantum bounce.\\

The results for the open models, both flat and hyperbolic geometries, are similar. This is because the expansion of the universe comes from the quantum bounce and continues. On the other hand, in contrast to the flat model, we consider the WEC and SEC for the matter entropy part here. These conditions yield the finding of the GSL regions where it is valid.

\subsection{GSL for a closed LQC model}
\label{subsec: GSL LQC k=+1}
In this section, we analyze the GSL for a closed universe in LQC using the modified Friedmann equations that, in the asymptotic limit, become the standard cosmic equations.\\
We use the results obtained from Section \ref{subsec: GSL f(A)}, where the GSL is defined by the $\dot{S}_g$ and $\dot{S}_m$ for $k=+1$, such that 

\begin{align}
    \label{k=-1 second law g LQC}
    \dot{S}_g&=-2\pi \dfrac{H\left(\dot{H}-\dfrac{1}{a^2}\right)}{\left(H^2+\dfrac{1}{a^2}\right)^2}\left[1+\alpha\left(H^2+\dfrac{1}{a^2}\right)\right],\\
        \label{k=-1 second law m}
    \dot{S}_m&=-\dfrac{4\pi (P+\rho)H}{T\left(H^2+\dfrac{1}{a^2}\right)^{5/2}}\left[\dot{H}+H^2\right]\nonumber\\
    &=-\dfrac{16\pi(P+\rho)H(\dot{H}+H^2)}{\left\vert2H^2+\dot{H}+\dfrac{1}{a^2}\right\vert\left(H^2+\dfrac{1}{a^2}\right)^2},
\end{align} 
with the apparent radius $R_A=1/\sqrt{H^2+\frac{1}{a^2}}$. To find the validity regions of the GSL, we consider points where $\dot{H}-1/a^2$ changes sign, which were found in Section \ref{sub: closed LQC model}. Furthermore, the WEC and the SEC are fulfilled, which are the conditions for standard matter. For the gravitational entropy, the possible values of the logarithmic correction are studied. With this in mind, we can analyze the different cases.\\

\textbf{Case $\alpha=0$}. For the case when there is no logarithmic correction, or equivalently, when $S_g=A_A/4$. If we assume that $\dot{H}-1/a^2\geq 0$, then $\dot{S}_g< 0$. On the other hand, the matter entropy  $\dot{S}_m<0$. Therefore, the GSL is not obeyed.\\
Now, if $\dot{H}-1/a^2< 0$, then $\dot{S}_g\geq 0$. For this assumption, we have two possibilities. First, when $\dot{H}+H^2\geq 0$, $\dot{S}_m<0$ is true. Thus, the GSL is valid only if $\dot{S}_g\geq \dot{S}_m$. Second, when $\dot{H}+H^2< 0$, then $\dot{S}_m>0$. Hence, $\dot{S}_T\geq 0$, the GSL is fulfilled.\\

\textbf{Case $1+\alpha(H^2-\frac{1}{a^2})\geq 0$}. This case includes at least the situation when $\alpha$ takes positive or zero values. Note that if $\dot{H}-1/a^2\geq 0$, we have that both $\dot{S}_g<0$ and $\dot{S}_m<0$. Thus, the GSL is not valid.\\
For the condition when $\dot{H}-1/a^2< 0$ is assumed, the gravitational part obeys $\dot{S}_g>0$. As previously considered, we have two options for the matter part. One of them, when $\dot{H}+H^2\geq 0$, yields $\dot{S}_m<0$. Therefore, the GSL is valid only if $\dot{S}_g\geq \dot{S}_m$. On the contrary, when $\dot{H}+H^2<0$, we have that the GSL is valid. These results are similar to those for the case when there is no logarithmic correction.\\

\textbf{Case $1+\alpha(H^2-\frac{1}{a^2})< 0$}. When it is assumed $\dot{H}-1/a^2\geq 0$, we have that $\dot{S}_g> 0$ and $\dot{S}_m<0$. Hence, the GSL is valid only if $\dot{S}_g\geq\dot{S}_m$.\\
Furthermore, when $\dot{H}-1/a^2< 0$ is considered, the gravitational part $\dot{S}_g<0$ and the matter part is analyzed in two options: i) when $\dot{H}+H^2\geq 0$, then $\dot{S}_m<0$. Therefore, the GSL is violated. ii) However, when $\dot{H}+H^2< 0$, we have $\dot{S}_m\geq 0$, and the GSL is valid only if $\dot{S}_m\geq \dot{S}_g$.\\

As for the $k=-1$ LQC model, the matter entropy part includes the WEC and SEC for the matter, it is not sufficient to consider the WEC as in the flat model.

Despite the results for the open and closed cases apparently being so similar, the regions where the quantity $\dot{H}-1/a^2$ changes sign are physically different because, in the closed case, there is a maximum size of the universe before the contraction until a quantum bounce occurs again. In comparison with the open cases, the universe grows infinitely after the first quantum bounce. The GSL is violated near to the both quantum bounces, and it is valid at large scales.

\section{An Extended Generalized Second Law}
\label{Appe: negative T}
In order to solve the violations of the GSL in LQC, it is possible to consider new avenues. Negative absolute temperature (NAT) systems have been studied, establishing specific conditions under which these systems can be obtained in statistical mechanics and verified experimentally. These conditions involve a bounded phase space and ensure that a NAT system is isolated from another with an absolute positive temperature (See more details in \cite{Ramsey-NegativeTemp:1956zz, BALDOVIN20211-NegativeTemp,Corichi:2025tts}). To explore them, we use a definition of temperature beyond kinetic energy as follows 
\begin{align}
    \label{temperature def}
    \dfrac{1}{T}=\left(\dfrac{\partial S}{\partial E}\right)_{X},
\end{align}
where $( \ \ )_X$ indicates that, for the partial differentiation, the variables $X$ should remain constant. These variables appear as additional differentials in the first law of thermodynamics, $E$ and $S$ are the energy and entropy of the system, respectively. 

The second law of thermodynamics states that the change in entropy always increases or remains equal when we consider positive temperatures for isolated systems. However, if we consider the possibility of obtaining a system with a negative absolute temperature it is natural to ask if the laws of thermodynamics remains without changes or we need to modify them. From the evidence of NAT system, we propose an extension of the second law, where its validity would arise from \cite{Corichi:2025tts} 
\begin{align}
\label{new second law}
    \mathrm{sign}(T)\mathrm{d}S\geq 0,
\end{align}
that is when $\mathrm{d}S\leq 0$ and $T<0$, or when $\mathrm{d}S\geq0$ with $T>0$. This extension includes the standard second law that is valid when we consider positive temperatures and the 'paradoxical' behavior that present the second law when a system admits NAT.

For the gravitational case, particularly the cosmological case, which is the case we have focused on throughout this article. We consider that the temperature is proportional to the surface gravity rather than the absolute value of surface gravity (\ref{cosmic temperature})
\begin{align}
    \label{negative temperature}
    T=\dfrac{\kappa}{2\pi}.
\end{align}
The universe in LQC fulfills the conditions to admit NAT, we can assign a temperature to each microstate, the phase space is bounded and  it is an example of a perfectly isolated system.

Due to the way to relate the temperature with a geometrical quantity is via surface gravity. Note that the surface gravity (\ref{surface gravity}) can take negative values when $\dot{H}+2H^2+\frac{k}{a^2}>0$ and positive values when $\dot{H}+2H^2+\frac{k}{a^2}<0$, this last is equivalent to using $T>0$. The value $T=0$ is not allowed because the second law of matter prohibits it in (\ref{f(A) second law m}).

In addition, in Section \ref{subsec: GSL in LQC}, we determine the validity regions of the standard GSL for positive temperatures. We now explore the conditions under which an \textit{Extended Generalized Second Law} (EGSL) is valid, taking into account eq.(\ref{negative temperature}) in (\ref{new second law}). By incorporating NAT into the overall description of the Universe. Since the temperature is proportional to the surface gravity (\ref{negative temperature}), different phases may arise in which gravity behaves attractively or repulsively. This distinction is relevant in cosmology for understanding the global evolution of the Universe.

\subsection{EGSL for a flat LQC model}

The conclusions in Section \ref{subsec: GSL in LQC}, when $\dot{H}+2H^2<0$, that is $T>0$, remain equal. Nevertheless, when we focus on the case $T<0$, namely, $\dot{H}+2H^2>0$. We note significant differences in the validation regions of the EGSL; for instance, for the flat LQC model with logarithmic corrections. 

\textbf{Case $\alpha=0$}. For $0<\rho<\rho_0$, 
\begin{itemize}
    \item when $\dot{H}+H^2>0$, implies that $\dot{H}+2H^2>0$ and $T<0$, the GSL is valid due to $\dot{S}_g>0$ and $\dot{S}_m>0$. However, under these conditions, the EGSL is not fulfilled.
    \item when $\dot{H}+H^2<0$. Here, there are two cases $2H^2+\dot{H}>0$ and $2H^2+\dot{H}<0$. For the latter option when $T>0$, both the GSL and the EGSL are agree and valid. However, when $2H^2+\dot{H}>0$ the EGSL is valid if $\dot{S}_g\geq \dot{S}_m$.
\end{itemize}

 For $\rho_c<\rho<\rho_0$, we have $\dot{H}>0$, then $\dot{H}+2H^2>0$, equivalently $T<0$. Thus, $\dot{S}_T\geq0$ is right if $\dot{S}_m\geq \dot{S}_g$. Contrary, the EGSL is valid when $\dot{S}_g\geq \dot{S}_m$. 
 
This same analysis for $\alpha=0$ obtains similar results for the case with positive values of $\alpha$.

\textbf{Case $\alpha=\alpha_0$}. For $0<\rho<\rho_0$ when $\dot{H}+H^2>0$ and $T<0$, the EGSL is invalid. In addition, when $\dot{H}+H^2<0$ and $\dot{H}+2H^2>0$, the GSL is valid is valid when  when $\dot{S}_m\geq\dot{S}_g$. In contrast, the EGSL is obeyed if $\dot{S}_g\geq\dot{S}_m$

For $\rho_0<\rho<\rho_c$, we have that $2H^2+\dot{H}>0$, which implies $T<0$. Here, the EGSL is fulfilled only when $\dot{S}_g\geq\dot{S}_m$.

The case when $\alpha_0<\alpha<0$ has the same results as for the case $\alpha=\alpha_0$, because $H^2(\rho_0)$ is the maximum.

\textbf{Case $\alpha<\alpha_0<0$}. For this case there exist two options when $i)$ $1+\alpha H^2\geq 0$ and $ii)$ $1+\alpha H^2< 0$, the option $i)$ implies the same results as in previous cases. However, the second option brings to different results:

When $\dot{H}\geq 0$, which implies that $T<0$. Therefore, $\dot{S}_T\geq 0$ and the EGSL is invalid under these conditions.

Now, if $\dot{H}<0$ and $\dot{H}+H^2>0$, then $T<0$. Therefore, the EGSL is valid only if $\dot{S}_g\geq\dot{S}_m$. Additionally, when $\dot{H}<0$ and $\dot{H}+H^2<0$, but $\dot{H}+2H^2>0$. Therefore, the EGSL is completely fulfilled. Finally, when $\dot{H}<0$ and $\dot{H}+H^2<0$, but $\dot{H}+2H^2<0$, that is $T>0$. Both the GSL and EGSL remain valid only if $\dot{S}_m\geq \dot{S}_g$.

We note that the EGSL is valid for every case just after the quantum bounce, in contrast with the GSL, which is violated. In addition, when $T<0$ and the GSL were valid, if the second law for the gravitational or the matter part dominated, with the EGSL, the dominated part changes. Finally, when $\dot{S}_g$ and $\dot{S}_m$ are both positive, the EGSL is violated at negative absolute temperatures.

\subsection{EGSL for an open LQC model}
\label{subsec: ESGSL k=-1}
In this section, the application of the extended generalized second law is studied in the $k=-1$ LQC model to address the violations of the GSL from Section \ref{Subsec: GSL k=-1}, where the results for positive absolute temperatures ($T>0$) were analyzed. Here, they are again valid. However, we focus on the NAT and the modification due to this proposed extension given by the EGSL.\\

\textbf{Case $1+\alpha(H^2-\frac{1}{a^2})\geq 0$}. This case includes when $\alpha$ is negligible. When $\dot{H}>0$, implies that $T<0$. Here, the EGSL is valid only if $\dot{S}_g\geq\dot{S}_m$. Also, $T<0$ when $\dot{H}+\dfrac{1}{a^2}<0$ and $\dot{H}+H^2>0$. Under these conditions the EGSL is not valid. On the other hand, when $\dot{H}+\dfrac{1}{a^2}<0$ and $\dot{H}+H^2<0$, there are two possibilities. In the first one, when $\dot{H}+2H^2-1/a^2>0$, the EGSL is valid only if $\dot{S}_m\geq\dot{S}_m$. Finally, when $\dot{H}+2H^2-1/a^2<0$, namely $T>0$, both the GSL and EGSL remain totally valid because $\dot{S}_g>0$ and $\dot{S}_m>0$.

\textbf{Case $1+\alpha(H^2-\frac{1}{a^2})<0$}. A similar anlysis is made as in previous case. We assume that $\dot{H}+1/a^2>0$, which implies that $T<0$. Thus, the EGSL is violated. Now, when $\dot{H}+1/a^2<0$ and $\dot{H}+H^2>0$, then $T<0$. Hence, the EGSL is fulfilled only if $\dot{S}_g\geq\dot{S}_m$. Finally, for $\dot{H}+1/a^2<0$ and $\dot{H}+H^2<0$, there are two options. The first option, when $T<0$, the GSL is valid while the EGSL is violated because both $\dot{S}_m>0$ and $\dot{S}_g>0$. The second option, when $T>0$, the results are the same both GSL and EGSL, that are valid when $\dot{S}_m\geq\dot{S}_g$.

The EGSL is valid in an extended region compared to the GSL; for instance, in the last case considered, it is valid for every possibility and extends the validation for the other cases even when we are close to the quantum bounce. This is a novel result that is consistent with the results for the EGSL in the flat case in LQC.

\subsection{EGSL for a closed LQC model}
\label{subsec: EGSL k=+1}
As for the $k=0,-1$ models in LQC, we consider the differences when we open the possibility of the NAT systems. The results when $T>0$ for the closed model in LQC are studied in Section \ref{subsec: GSL LQC k=+1} remain valid for this discussion. In this section, we only include the new results when the NAT is considered and the extension of the second law is implemented.\\

\textbf{Case $1+\alpha(H^2-\frac{1}{a^2})\geq 0$}. This case includes the results when the logarithmic correction is not considered; that is $\alpha=0$. In order to open the possibility of obtaining NAT in our system, we analyze the different possibilities. The first consideration using the SEC is given when $\dot{H}+1/a^2>0$, which implies that $T<0$. Here, the EGSL is fulfilled only if $\dot{S}_g\geq \dot{S}_m$. Now, when $\dot{H}+1/a^2<0$ and $\dot{H}+H^2>0$ yields $T<0$. Thus, the GSL is valid while the EGSL is not obeyed due to $\dot{S}_g>0$ and $\dot{S}_m>0$. Furthermore, if $\dot{H}+1/a^2<0$ and $\dot{H}+H^2<0$, there are two options: $i)$ when $\dot{H}+2H^2-1/a^2>0$, the EGSL is valid if $\dot{S}_m\geq\dot{S}_g$. Finally, $ii)$ when $\dot{H}+2H^2-1/a^2<0$, we obtain $\dot{S}_m>0$ and $\dot{S}_g>0$. Hence, the GSL and the ESGL are valid and coincide because $T>0$.    \\

\textbf{Case $1+\alpha(H^2-\frac{1}{a^2})<0$}. Among the different possibilities, we start considering when $\dot{H}+1/a^2>0$, which leads to $T<0$. Note that both $\dot{S}_m>0$ and $\dot{S}_g$. Therefore, the EGSL is violated while the GSL is valid when $\dot{S}_g\geq\dot{S}_m$. Now, we assume that $\dot{H}+1/a^2<0$ and $\dot{H}+H^2>0$, which implies that $T<0$. Hence, the EGSL is valid only if $\dot{S}_g\geq\dot{S}_m$. Under the assumption that $\dot{H}+1/a^2<0$ and $\dot{H}+H^2<0$, there are two options. The first implication yields obtaining $T<0$, here both $\dot{S}_g<0$ and $\dot{S}_m<0$. Thus, the EGSL is fully valid. The second implication leads to $T>0$, which implies that the GSL and EGSL remain valid when $\dot{S}_m\geq\dot{S}_g$.    \\

The validity regions of the EGSL are similar to those in the open LQC models. However, for the closed model, there are more regions where $\dot{H}-1/a^2$ changes sign, not only near the quantum bounce. The Universe expands until it contracts again. In these regions, the temperature also changes sign, and the EGSL is implemented when we open the option to have negative absolute temperatures. Similarly to the previous examples in LQC, we note that this new way of viewing the second law of thermodynamics helps us extend the regions of validation due to NAT.

\section{Time Arrow}
\label{Time arrow}
In previous sections, the thermodynamic laws in loop quantum cosmology were analyzed for a single cosmological branch. However, it is natural to inquire about the behavior preceding the quantum bounce and whether any remnants of this phase could be detected after the bounce \cite{Corichi:2007am}. For the LQC models considered here, the two branches connected by the bounce are symmetric. It is straightforward to verify that the effective dynamical equations (\ref{eff friedmann eq})–(\ref{eff dot H}) are invariant under time reversal. The surface gravity (\ref{FLRW surface gravity}) is likewise invariant, since the apparent radius (\ref{apparent radius}) is time-reversal symmetric. As a result, even in the presence of negative absolute temperature (NAT), the temperature definition (\ref{negative temperature}) is invariant under this transformation. Nevertheless, while the thermodynamic framework of LQC appears symmetric, the gravitational entropy variation is not time-reversal invariant, in contrast to the matter entropy, which is invariant, as can be seen from (\ref{f(A) second law g}) and (\ref{f(A) second law m}), respectively.

From these results, we conclude that in open LQC models, for both cosmological branches, the temperature is negative in the vicinity of the quantum bounce and positive far from it. Three transition points are identified: two associated with sign changes of the temperature and one corresponding to the transition between branches. The non-invariance of the gravitational entropy variation under time reversal provides a natural arrow of time consistent with the second law.

By contrast, in the closed LQC model, an additional sign change of the temperature occurs. Near the point of maximum expansion, the temperature remains positive on both sides of the maximum. Subsequently, the Universe recollapses until a new quantum bounce is reached, where the temperature again becomes negative.

The expansion and contraction of the universe are closely linked to gravitational entropy. The GSL in closed cosmological models has been an important topic since classical cosmology, where the thermodynamic arrow of time derived from the second law was already discussed \cite{Hawking:1985af}. This issue remains significant in quantum cosmology, where quantum effects resolve classical singularities and connect two cosmological branches via one or more quantum bounces. In this framework, the arrow of time can be defined through the entropy evolution dictated by the second law of thermodynamics.

\section{Discussion and Conclusions.}
\label{section:Conclusion}
In summary, this article examined the GFL and the distinctions between the GSL and EGSL within effective models of LQC, considering flat, open, and closed geometries. This analysis completes the thermodynamic investigation of all possible geometrical configurations in loop quantum cosmology, incorporating logarithmic corrections to the gravitational entropy associated with the apparent horizon. Furthermore, the validity regions of the second law were determined for both positive and negative temperatures, together with their relation to the arrow of time.

The generalized first and second laws were analyzed by expressing entropy as a function of the apparent horizon area, a choice motivated by quantum corrections to black hole entropy obtained from state counting \cite{Ghosh:2004rq,Ghosh:2006ph,Engle:2009vc,Agullo:2010zz-entropyBH,Ashtekar:1997yu,Rovelli:1996dv,Hossenfelder:2012tc,Meissner:2004ju,Domagala:2004jt}. We further examined alternative quantum cosmological models that can be recast in standard cosmological form, enabling the application of conventional thermodynamic methods \cite{Hayward:1997jp,Hayward:1998ee,Cai:2005ra,Faraoni:2015ula}. In this approach, rewriting the effective cosmological equations requires determining the effective energy density and the effective pressure of matter. We obtained general conditions for the validity of the gravitational and matter second laws, as well as the GSL. However, these conditions depended on the effective model with which we worked and the function (of area) that we chose for entropy.

Logarithmic entropy corrections arising from black hole microstate counting in LQG \cite{Ghosh:2004rq,Ghosh:2006ph,Agullo:2010zz-entropyBH,Meissner:2004ju,Engle:2009vc,Domagala:2004jt,Ashtekar:1997yu,Rovelli:1996dv} were incorporated in the study of the GFL and GSL for a broad class of quantum cosmological models. In particular, effective LQC models with spatial curvature were reformulated into the standard cosmological framework. While flat effective LQC scenarios have been previously studied \cite{Sadjadi:2012wg,Zhang-Thermo:2021umq,Li-Thermo:2008tc,Silva:2023ent}, we extended these analyzes to include nonzero curvature cases, thereby completing earlier investigations of the GFL and GSL in effective LQC models. We also obtained the conditions under which the GSL is satisfied for arbitrary values of the logarithmic correction parameter $\tilde{\alpha}$. 

The GFL was found using entropy as a function of apparent area and temperature, which is proportional to the absolute value of the surface gravity, and then adding the second law for the matter part, which is ordinary matter that obeys the WEC and the SEC. For the flat LQC model, it is sufficient to assume the WEC; for the LQC models with spatial curvature distinct from zero, both conditions are assumed. In particular, if we consider logarithmic corrections, these affect the total energy, total entropy, total work, and the apparent volume (\ref{GFL of LQC}). These logarithmic corrections are negligible when the apparent area is large or when $\alpha=0$. In addition, for LQC with $ \alpha=0$ \cite{Zhang-Thermo:2021umq,Li-Thermo:2008tc}, we found that quantum corrections were small when $\rho_*>>\rho$, and we recovered the standard GFL in cosmology.

The analysis for the GSL of an effective flat LQC model \cite{Ashtekar:2006wn,Ashtekar:2007em} with logarithmic corrections looks for validity conditions for $\tilde{\alpha}=0$, $\tilde{\alpha}>0$ \cite{Hod:2004di}, $\tilde{\alpha}_0$ \cite{Sadjadi:2012wg}, $\tilde{\alpha}_0<\tilde{\alpha}<0$ \cite{Ghosh:2004rq,Ghosh:2006ph,Agullo:2010zz-entropyBH,Meissner:2004ju,Engle:2009vc,Domagala:2004jt,Ashtekar:1997yu,Rovelli:1996dv}, and $\tilde{\alpha}<\tilde{\alpha}_0<0$. Particularly for the flat LQC model, we focused on the effective Raychaudhuri equation that changes sign at $\rho_0=\rho_c/2$, and we could equivalently study the state equation (\ref{GSL state eq}) for $P=w\rho$, where we found that just after the quantum bounce, the GSL was invalid; only when $\tilde{\alpha}<\tilde{\alpha}_0<0$ and $H^2(\rho_2)<H^2(\rho)<H^2(\rho_0)$ could it be valid. After the transition point $\rho=\rho_0$ or at $\rho_0<\rho<\rho_c$, the GSL is valid even though the second law of gravity or the matter part fails; only for $\tilde{\alpha}<\tilde{\alpha}_0<0$, $H^2(\rho_2)<H^2(\rho)<H^2(\rho_0)$, and $H^2+\dot{H}>0$ the GSL not hold in this region.

For open LQC models ($k=0,-1$), the regions of validity of the generalized second law (GSL) are physically equivalent: the GSL is violated near the quantum bounce and satisfied at late times, as both models describe universes expanding from the bounce toward infinity. By contrast, the effective closed LQC model exhibits a cyclic behavior, expanding from an initial bounce and later recollapsing toward a second bounce. In this scenario, the validity of the GSL includes the point of maximum expansion as well as the additional bounce in the contracting phase. 

We studied the possibility of negative absolute temperature (NAT), which is considered  proportinal to the surface gravity (\ref{negative temperature}). An extension of the generalized second law (97) was introduced, which remains valid both in regimes where the standard GSL holds for $T>0$ and in regimes where the standard GSL fails but $T<0$. This extended generalized second law (EGSL) enlarges the domain of validity for the flat LQC model with logarithmic corrections, compared to the standard analysis restricted to $T>0$. In particular, the EGSL holds immediately after the quantum bounce, unlike the standard GSL.

The temperature at the horizon is related to the surface gravity. In the standard formalism, only positive temperatures are considered, corresponding to an attractive gravitational interaction. However, by allowing negative temperatures in the description, one may interpret them as different gravitational phases in which gravity becomes either attractive or repulsive. This idea is particularly relevant in cosmology, where different regions of the universe may undergo expansion or collapse.

A standard interpretation of the NAT \cite{Ramsey-NegativeTemp:1956zz,BALDOVIN20211-NegativeTemp,Corichi:2025tts} comes at the transition point between positive and negative temperatures, and the NAT is 'hotter' than positive temperature; this is the reason why we only observe positive temperatures. This may explain why only positive temperatures are observed in nature. Such an interpretation is also compatible with cosmological scenarios for the origin of the Universe, where regions near the quantum bounce are expected to be significantly 'hotter' than large-scale regions.

The thermodynamics of bounce in LQC was explored through the time-reversal transformation. It was shown that the temperature is negative absolute in the vicinity of the quantum bounce and positive absolute far from it. Moreover, effective open LQC models exhibit three transition points: two points, due to the temperature changes in sign, one in each branch, and one additional point that corresponds to the quantum bounce, while the closed model contains an additional transition point. Although the effective dynamics of LQC are modified, the temperature and $\dot{S}_m$ remain invariant under time reversal. However, since $\dot{S}_g$ is not invariant under this symmetry, the second law naturally defines a preferred arrow of time.

The apparent horizon depends only on the metric, and it is independent of the gravity theory with which we work \cite{Cai:2005ra,Hayward:1997jp,Hayward:1998ee,Faraoni:2015ula}. However, we could choose and explore the thermodynamics using another horizon. For example, in \cite{Ashtekar-Ewing:2008gn} the horizon of particles helped to solve the covariant entropy conjecture \cite{Bousso:1999xy} where the entropy was of the order of $R^3$ instead of $R^2$. Another example in which the chosen horizon is relevant to the results. From (\ref{first law matter}), it was possible to derive the first law of matter (\ref{dot Sm}) using the apparent volume $V_A$ for $k=0$. In comparison, if we use the physical volume $V$, the second law of matter yields $\dot{S}_m=0$ because of $H=\frac{\dot{V}}{3V}$ instead of $-\frac{\dot{H}}{H}=\frac{\dot{V}_A}{3V_A}$. Apparent and physical horizons are equal only when $\dot{H}+H^2=0$, that is, in a static universe. Therefore, the thermodynamics in cosmology depends on the choice of the horizon \cite{Faraoni:2015ula}.

Although the thermodynamics of effective LQC models with spatial curvature $k=0,\pm 1$ have been investigated, alternative LQC formulations involve ambiguities associated with both classical and quantum descriptions, leading to notable differences in the effective dynamics \cite{Assanioussi:2018hee,Assanioussi:2019iye,Yang:2009fp,Yang:2022aec,Zhang:2021zfp,Gallegos:2026,Gallegos:2024ajc}. For instance, incorporating the Lorentzian term in loop quantization produces an emergent cosmological constant in the effective theory. The thermodynamic analysis of these alternative models, for both open and closed universes, is left for future work. The procedure of rewriting the effective equations in the standard cosmological form may then be used to identify the validity regions of the second law.

This article investigated cosmic thermodynamics in the semi-classical regime in the presence of the apparent horizon. The validity conditions may be modified if quantum entropy, rather than effective entropy, is considered in the cosmological context. Our experience dictates that the quantum description can solve (semi) classical problems. Nevertheless, a comprehensive analysis is deferred to future work.

The value of the Barbero-Immirzi parameter comes from the microstates counting of black holes in LQG; there are different methods and counting \cite{Ghosh:2004rq,Ghosh:2006ph,Agullo:2010zz-entropyBH,Meissner:2004ju,Engle:2009vc,Domagala:2004jt,Ashtekar:1997yu,Rovelli:1996dv}. The Barbero-Immirzi parameter does not depend on the logarithmic correction because it takes an asymptotic limit when the area is large and logarithmic corrections are not relevant. Therefore, the factor $\tilde{\alpha}_0$ \cite{Sadjadi:2012wg} would not affect the known values of the Barbero-Immirzi parameter.

Finally, the conditions for the violation of the GSL depend on the specific effective model and on the assumed entropy–area relation. It is therefore of interest to consider alternative effective models in LQC or other gravitational frameworks, as well as different entropy ansatz as functions of the apparent horizon area \cite{Silva:2023ent}, to further analyze the quantum bounce using the methods developed in this work. 

\section{Acknowledgments.}
\label{section:Acknowledgments}
O.G. thanks that this work was supported by UNAM Posdoctoral Program (POSDOC). Additionally, the work of O.G. was financially supported by SECIHTI postdoctoral fellowships and SNII with CVU 786532.

\bibliography{ref}

\end{document}